\documentclass[pra,superscriptaddress,epsfigure,twocolumn,showpacs,preprintnumbers,amsmath,amssymb]{revtex4-1}

\usepackage{graphicx}
\usepackage{color}
\usepackage{dcolumn}
\usepackage{bm}

\begin{document}

\title{Non-monotonic Casimir interaction: The role of amplifying dielectrics}

\author{Morteza Soltani}
\address{Department of Physics, University of Isfahan, Isfahan 81746, Iran}

\author{Jalal Sarabadani}
\email{jalal.sarabadani@aalto.fi}
\address{Department of Applied Physics, and COMP Center of Excellence,
    Aalto University School of Science, P.O. Box 11000, FI-00076 Aalto, Espoo, Finland}

\author{S. Peyman Zakeri}
\address{Department of Physics, University of Isfahan, Isfahan 81746, Iran}

\begin{abstract}
The normal and the lateral Casimir interactions between corrugated ideal metallic plates in the presence 
of an amplifying or an absorptive dielectric slab has been studied by the path-integral quantization technique. 
The effect of the amplifying slab, which is located between corrugated conductors, is to increase the 
normal and lateral Casimir interactions, while the presence of the absorptive slab diminishes the interactions. 
These effects are more pronounced if the thickness of the slab increases, and also if the slab comes closer 
to one of the bounding conductors. When both bounding ideal conductors are flat, the normal Casimir force 
is non-monotonic in the presence of the amplifying slab and the system has a stable mechanical equilibrium state, while 
the force is attractive and is weakened by intervening the absorptive dielectric slab in the cavity. 
By replacing one of the flat conductors by a flat ideal permeable plate the force becomes non-monotonic 
and the system has an unstable mechanical equilibrium state in the presence of either an amplifying or an absorptive slab. 
When the left side plate is conductor and the right one is permeable, the force is non-monotonic in the 
presence of a double-layer DA (dissipative-amplifying) dielectric slab with a stable mechanical equilibrium state, 
while it is purely repulsive in the presence of a double-layer AD (amplifying-dissipative) dielectric slab.
\end{abstract}

\pacs{42.50.Lc, 03.70.+k, 42.50.-p}

\maketitle

\section{Introduction}\label{intro}

An interesting macroscopic result of confinement of the quantum electromagnetic (EM) field
between two ideal parallel flat conductors is the Casimir effect \cite{Casimir}.
Comparing the energy of the system in the presence and in the absence of bounding ideal metallic plates
leads to a finite attractive potential energy. The derivation of this finite energy with respect
to the distance between two plates, $H$, is the Casimir force, $F_{\textrm{C}}$. The Casimir force per unit
area, $S$, is then read as $\frac{{F_{\textrm{C}} }}{S} =  - \frac{{\hbar c\pi ^2 }}{{240H^4 }}$,
where $\hbar$ is the Planck constant multiplied by $\frac{1}{2\pi}$, and $c$ is the speed of light in the
vacuum \cite{MilonniCook}.
The magnitude of this force is significant when $H$ is less than a micron
\cite{Lambrecht_Phys_World}. Therefore the Casimir effect must be taken into account in
designing micro- and nano-devices \cite{Chan_Capasso,Buks_Rukes,Broer,Nasiri_Appl_Phys_Lett}.

\begin{figure}[b]
\includegraphics[scale=0.2]{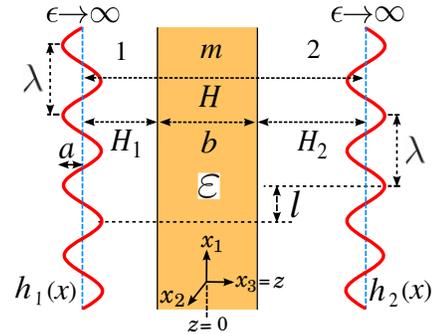}
\caption{(Color online) The schematic figure of a dielectric slab with dielectric function $\varepsilon$
and thickness $b$ immersed in the quantum vacuum and enclosed by two rough metallic plates. The mean
distance between two plates is $H$. The corrugation amplitude and wavelength for both plates are $a$
and $\lambda$, respectively. The mean distance between the left metallic plate and the left side of the
dielectric slab is $H_1$, while the mean distance between the right metallic plate and the right side of
the dielectric slab is $H_2$. The lateral mismatch between metallic plates is $l$. The corrugations on
the left and on the right conductors are given by two height functions $h_1 (x)$ and $h_2 (x)$, respectively.
The regions labeled by 1 and 2 are filled by the quantum vacuum while the dielectric slab is labeled by $m$.}
\label{fig:schematic_1}
\end{figure}

In addition to the normal Casimir force, corrugated conductors can experience lateral Casimir force
due to the translational symmetry breaking which has been observed experimentally in 2002 \cite{Mohideen_lateral_PRL_PRA}.
The lateral Casimir effect between two rough conductors with sinusoidal corrugation has been studied
theoretically in the context of path-integral formalism \cite{Golestanian_PRA_1998,Golestanian_PRA_2003,Jalal_PRA_2006}.
This force is very sensitive to the characteristics of the rough surfaces and also sensitive to the medium in between them.
For a system composed of two rough ideal conductors immersed in the quantum vacuum, at fixed mean separation distance
between two conductors, $H$, with $\lambda$ as the wavelength of corrugations on both plates, the stable and unstable
equilibrium states for the position of the rough plates in the lateral direction occur when $l= (n+\frac{1}{2})\lambda$ and
$l= n \lambda$, respectively \cite{Golestanian_PRA_2003}, whereas for the system composed of a rough ideal conductor and
an infinitely permeable corrugated plate, with the same $H$ and $\lambda$, the stable and unstable equilibrium states for the
position of the plates in the lateral direction occur when $l= n \lambda$ and $l= (n+\frac{1}{2})\lambda$, respectively
\cite{kiani_PRA_2012}. 

Moreover, the Casimir force also strongly depends on the characteristics of the medium
between the bounding plates. In a series of papers, it has been shown that the presence of absorptive dielectric medium 
between two bounding conductors can weaken the normal and lateral Casimir interactions 
\cite{Soltani_flat_PRA_2010,Soltani_rough_PRA_2010,Soltani_annals_2011,Zakeri_PRA_2012}.
The reason is that the absorptive dielectrics can dissipate the EM energy \cite{welsch1,welsch2,sutorp}.
In the other hand, it has been shown that if the energy is pumped into the medium artificially, for example by lasers,
the EM energy can be amplified in some regions of frequency 
\cite{welsch-casimir,ehsan1, Matloob_PRA_2001, Matloob_PRA_1997, Welsch_EPJST_2008}.
These kinds of media are called {\it amplifying} media, where the EM energy can be amplified. Although 
the physical properties of these media are different from those of dissipative dielectrics, the same formal methods can be 
used to quantize the EM field in the presence of amplifying dielectrics as of dissipative dielectrics
\cite{Matloob_PRA_2001, Matloob_PRA_1997, Welsch_EPJST_2008}.

To study an amplifying medium, one can introduce a susceptibility which must satisfy the Kramers-Kronig relations.
Similar to the dissipative (or absorptive) dielectrics, the dielectric function of an amplifying medium, that is
$\varepsilon_{\textrm{amp}} (\omega)$,
should have an imaginary part. But for dissipative dielectrics, the imaginary part of the dielectric function is positive,
i.e. $\varepsilon_{{\textrm{I,disp}}} (\omega)>0$, whereas for amplifying media, the imaginary part is negative, i.e.
$\varepsilon_{{\textrm{I,amp}}} (\omega)<0$ \cite{Matloob_PRA_1997}. The same is true for the imaginary parts of the
permeability of the dissipative and amplifying media \cite{ehsan1}. One can quantize the EM field in the presence of
the amplifying medium by adding a noise term into Maxwell equations \cite{Matloob_PRA_1997, Welsch_EPJST_2008, Welsch_PRA_2000},
or using the canonical approach \cite{Matloob_PRA_1997}. For quantization of the EM field in the presence of an amplifying
medium, we use the path-integral formalism \cite{Soltani_flat_PRA_2010}. It should be noted that although the methods
for field quantization which we use here for the amplifying and dissipative systems are the same, however the characteristics
of these two kinds of systems are completely different. For example, vacuum fluctuations in the presence of dissipative and 
amplifying media behave in different ways. Moreover, an amplifying medium can be used to compensate the effect of dissipation
\cite{Matloob_PRA_1997}.

In the present paper, we study and compare the normal and the lateral Casimir interactions between two rough ideal conductors
in the presence of an intervening dissipative or amplifying slab, or a double-layer with mixture of them. This paper is organized
as follows:
In Sec. \ref{formalism} we generally introduce the path-integral formalism to obtain the Casimir interaction energy for 
the system composed of a dielectric slab between two ideal rough metallic plates. Section \ref{sec_flat} is devoted to study the 
Casimir interaction in a system composed of either an absorptive or an amplifying slab or a double-layer dielectric slab between 
two flat bounding ideal plates. In Sec. \ref{sec_corrugation} we calculate and compare the normal and the lateral Casimir 
interactions due to the corrugation on the bounding plates in the presence of either an absorptive or an amplifying slab. 
Finally we wrap up the paper with a conclusion in Sec. \ref{sec_conclusion}.

\section{Formalism}\label{formalism}
\subsection{Quantization of the EM field}\label{quantization}

In this section we investigate the Casimir interaction between two ideal corrugated conductive plates in the
presence of either a dissipative or an amplifying slab as depicted in Fig.~\!\ref{fig:schematic_1}. $X= (x_0, \bf{x})$ is
the coordinate of a point in four-dimensional time-space with the temporal coordinate $x_0$, and the spatial coordinates
${\bf{x}} = (x_1, x_2 , x_3)$, $x= (x_1, x_2)$. $z$ component of $X_{\textrm{1}}$, which is in the direction of $x_3$,
on the left conductor is $z= -H_1 - b/2 +h_1(x)$, while $z$ component of $X_{\textrm{2}}$ on the right conductor is 
$z= H_2 +b/2 + h_2(x)$,
where $h_1 (x)$ and $h_2 (x)$ are the deformation fields on the left and on the right ideal plates, respectively. 
The average distance between
two conductors is $H$, the corrugation amplitude for both conductors is $a$, the wave length of the corrugation in $x_1$
direction for both conductors is $\lambda$, $l$ is the lateral shift between surfaces in $x_1$ direction, $H_1$ is
the mean distance of the left side of the slab and the left conductor, $H_2$ is the mean distance of the right side
of the slab and the right conductor, and $b$ is the thickness of the slab.

To study the Casimir interaction between two corrugated ideal metallic plates in the presence of an
intervening dissipative or amplifying slab (see Fig.~\!\ref{fig:schematic_1}), we first use the path-integral method
to quantize the EM field. As we are only considering a uni-axial cavity, therefore the EM
field can be decomposed into the transverse magnetic (TM) and transverse electric (TE) waves which
satisfy Dirichlet (D) and Neumann (N) boundary conditions (BCs), respectively on both plates
\cite{Golestanian_PRA_2003, Jackson, Jalal_PRA_2006}. The D BC should be satisfied by
TM waves, that is $\phi(X_{\textrm{a}})= 0$, on both plates, while TE waves satisfy the N BC, that is
$\partial_n \phi(X_{\textrm{a}})= 0$, where $\textrm{a}= 1,2$ determines the left and right plates, respectively.
Consequently, after decomposing the EM field into TM and TE waves we quantize a massless scalar Klein-Gordon field,
$\phi$, in the presence of a dielectric slab between two conducting plates. To this end, we use the same method
as of Refs.~\!\cite{Zakeri_PRA_2012, Soltani_flat_PRA_2010, Soltani_annals_2011, Soltani_rough_PRA_2010}.
The dielectric medium can be a dissipative dielectric slab with a positive imaginary part of its
dielectric function, that is $\varepsilon_{{\textrm{I,disp}}}(\omega) > 0$, or it can be an amplifying slab with
a negative imaginary part of its dielectric function in some frequency ranges, that is
$\varepsilon_{{\textrm{I,amp}}}(\omega) < 0$. The dielectric medium can be a layered dielectrics with mixture of dissipative
and amplifying layers. Here, for the sake of simplicity and brevity we just present the formalism for
a slab with only one dielectric layer, that can be either an absorptive or an amplifying, in between the bounding conductors.
To extend the path-integral formalism to study the Casimir effect in the presence of an absorptive 
\cite{Soltani_flat_PRA_2010} or an amplifying medium \cite{Amooghorban_PRA_2011}, one must write the Lagrangian of the 
system with appropriate terms that leads to the correct equations of motion. To this end, we consider the following 
Lagrangian density 
\begin{equation}\label{total_Lagrangian}
{\cal L} ={\cal L}_{\textrm{sys}}  + {\cal L}_{\textrm{mat}}  + {\cal L}_{{\mathop{\rm \textrm{int}}} },
\end{equation}
where ${\cal L}_{\textrm{sys}}= \frac{1}{2}\partial ^\alpha  \phi (X)\partial _\alpha  \phi (X)$ is
the Lagrangian density of Klein-Gordon field, the summation over $\alpha= x_0, x_1, x_2 ~ {\textrm{and}} ~ x_3$ 
is assumed, and as explained in the above, $X = (x_0, \bf{x})$ is a coordinates of a point in four-dimensional time-space.
${\cal L}_{\textrm{mat}}  = \int\limits_0^\infty  d \omega ~ {\textrm{sgn}}
[\varepsilon_{\textrm{I}}(\omega)](\frac{1}{2} \dot{Y}_\omega^2 - \frac{1}{2} \omega ^2 Y_\omega ^2)$
is Lagrangian density of a matter field, where the matter is modeled by a continuum oscillator's field $Y_{\omega}$ 
\cite{Huttner_EPL_1992}, $\varepsilon_{\textrm{I}}(\omega)$ is the imaginary part of the dielectric function,
and ${\cal L}_{{\mathop{\rm \textrm{int}}} } = \phi \dot P$ is the interaction Lagrangian density between matter 
and scalar fields. $P = \int {d\omega \nu (\omega ) Y_\omega  }$ is the polarization of the medium, and $\nu (\omega)$ 
is the coupling function between the scalar and the matter fields \cite{Soltani_flat_PRA_2010}.

Using the Lagrangian in Eq.~\!(\ref{total_Lagrangian}), the conjugate of the canonical momentum
can be found. By employing the equal time commutation relations, one can quantize the field
and obtain the Hamiltonian of the system, and show that the scalar field $\phi$ satisfies the
following equation
\begin{equation}\label{varphi_relation}
(\nabla^{2}-\varepsilon (\omega)\omega ^2)\phi^{+}(\omega) = j^{+}_{N}(r,\omega).
\end{equation}
The positive frequency part of the current density operator, $j^{+}_{N}(r,\omega)$, is
\begin{eqnarray}\label{current_density}
j^{+}_{N}(r,\omega) &=& \Theta[\varepsilon_{\textrm{I}}(\omega)]
\sqrt{\frac{2\omega\varepsilon_{0}|\varepsilon_{\textrm{I}}(\omega)|}{\pi}} \hat{a}_{N}(r,\omega)\nonumber\\
&+& \Theta[-\varepsilon_{\textrm{I}}(\omega)]\sqrt{\frac{2\omega\varepsilon_{0}|\varepsilon_{\textrm{I}}(\omega)|}{\pi}}
\hat{a}^{\dag}_{N}(r,\omega),
\end{eqnarray}
where $\Theta(\ldots)$ is the Heaviside step function, $\hat{a}$ is a bosonic field with
bosonic commutation relation, and $\hat{a}^{\dag}$ is the complex transpose of $\hat{a}$.
Then, the Hamiltonian can be written as
\begin{equation}\label{Hamiltonian}
H=\int {d\omega \int {d^{3}r \sin[\varepsilon_{\textrm{I}}(\omega)] \hbar \omega \hat{a}^{\dag}_{N}(r,\omega)
\hat{a}_{N}(r,\omega) } },
\end{equation}
where
$\varepsilon_{\textrm{I}}(\omega)=1+\int {d\omega' \frac{\textrm{sgn}[\varepsilon_{\textrm{I}}(\omega')]
|\nu^{2}(\omega')|}{\omega^{2}-\omega'^{2}+\imath 0^{+}}  }$
leads to $\nu^{2}(\omega)= {\textrm{sgn}}[\varepsilon_{\textrm{I}}(\omega)]
\textrm{Im}[\varepsilon_{\textrm{I}}(\omega)]$.
The main difference between the calculations for the amplifying medium and dissipative one is the presence of
$\Theta[\varepsilon_{\textrm{I}}(\omega)]$ and $\textrm{sgn}[\varepsilon_{\textrm{I}}(\omega)]$ in the
above equations that have been
discussed extensively in \cite{Matloob_PRA_1997, welsch-casimir, Welsch_EPJST_2008, Welsch_PRA_2000}.
It should be mentioned that the Hamiltonian in Eq.~\!(\ref{Hamiltonian}) is consistent with the result of
Ref.~\!\cite{Welsch_EPJST_2008}.

To quantize the field, we calculate the generating function for the interacting system \cite{Ryder_book}.
The generating function is defined as
\begin{eqnarray}\label{generating_function}
{\cal Z} [J_\phi  ,J_P  ] &=& Z_0^{-1} e^{\imath \int dY {\cal L}_{{\mathop{\rm \textrm{int}}} }
[\frac{\delta}{\delta J_{\phi} (Y) }, \frac{\delta}{\delta J_{\omega} (Y) } ] } {\cal Z}_0 [J_\phi ,J_\omega ] \nonumber\\
&=& Z_0^{-1} \sum_{n=0}^{\infty} \frac{1}{n!} \bigg[ \imath   \int dY \frac{\delta}{\delta J_{\phi} (Y) }
\frac{\partial}{\partial y_0 } \frac{\delta}{\delta J_{P} (Y) }  \bigg]^n \nonumber\\
&\times& {\cal Z}_0 [J_\phi , J_\omega ] ,
\end{eqnarray}
where $Z_0= \int {\cal D} [\phi] \textrm{exp} \big( \imath \int dX {\cal L}_{\textrm{sys}} \big)$
is the generating function for free space,
${\cal Z}_{0} [J_\phi  ,J_\omega  ] = \int {{\cal D}[\phi ]} \prod\limits_\omega
{{\cal D}[Y_\omega ]} \exp \left\{ {\imath \int {dX\left[ {{\cal L}_{\textrm{sys}} +
{\cal L}_{\textrm{mat}}} \right.} } \right.
+ J_\phi  \phi  + \int {d\omega } J_\omega  \left. {\left. {Y_\omega  } \right]} \right\}$
is the generating function for noninteracting part of the system, and $J_{\omega}$ and $J_{\phi}$
are source fields that are coupled to the free fields $Y_{\omega}$ and $\phi$, respectively.
Then by integrating over $\phi$ and $Y_m$ the generating function can be written as
\begin{eqnarray}\label{completted generating_function}
{\cal Z} [J_{\phi}  ,J_P ] &=& \exp \bigg\{ \imath \int dX \int dX' \bigg[
J_{\phi}  G_{\phi ,\phi } (X - X') J_{\phi} \nonumber\\
&& \hspace{-1.5cm} +  J_{\phi}  G_{P,\phi } (X - X') J_P
+J_P G_{P,P} (X - X') J_P \bigg]  \bigg\},
\end{eqnarray}
where the Fourier transformation of the Green's function, $G_{\phi , \phi }$, is
\begin{equation}
G_{\phi ,\phi } ( q_0 , {\bf q} ) = \bigg[ {\bf q}^2  - q_0^2  \bigg(1+ \int d q'_0
\frac{ \textrm{sgn}[\varepsilon_{\textrm{I}}(q'_0)] \nu^2 (q'_0) }{ q^{'2}_0 - q_0^2 + \imath 0^{+} } \bigg) \bigg]^{-1},
\label{green_function_phi-phi}
\end{equation}
and for the other Green's functions can be calculated as
\begin{eqnarray}
G_{\phi ,P} (q_{0}, {\bf q} ) &=& \imath q_{0} \int d q'_0
\frac{ \textrm{sgn}[\varepsilon_{\textrm{I}}(q'_0)]   \nu^2 (q'_0) } { q^{'2}_0  - q_0^2
+ \imath   0^{+}  }G_{\phi ,\phi } (q_{0}, {\bf q} ), \nonumber\\
G_{P,P} (q_{0}, {\bf q} ) &=& \bigg[\int d q'_0 
\frac{ \textrm{sgn}[\varepsilon_{\textrm{I}}(q'_0)] \nu^2 (q'_0) } { q^{'2}_0 - q_0^2
+ \imath   0^{+}  } \bigg]^{2}G_{\phi, \phi } (q_{0}, {\bf q} ) \nonumber \\
&& + \frac{{\nu^2 (q_{0} )}}{q_{0} } + q_{0} ^2 ,
\label{green_function_otther}
\end{eqnarray}
where $\vec{q}= (q_0, {\bf q})$, ${\bf q}= (q_1, q_2)$ and $q_0$ is the temporal Fourier component.
The above Green's functions are fully in agreement with those in Refs.~\!\cite{Huttner_PRA_1992,Matloob_PRA_1996}.
Using the definition $\chi (q_{0})=  \int d q'_0 \frac{ \textrm{sgn}[\varepsilon_{\textrm{I}}(q'_0)]
\nu^2 (q'_0) } { q^{'2}_0  - q_0^2  + \imath   0^{+}  }$,
it can be shown that $G_{\phi,\phi}$ is the Green's function of Eq.~\!(\ref{varphi_relation})
and together with using the residues theorem, one can show that
$\textrm{Im}\chi (q_0)= \pi \textrm{sgn}[\varepsilon_{\textrm{I}}(q_0)] \nu^2 (q_0) /( 2 q_0)$.
Combining $\textrm{Im}\chi (q_0)$ with the definitions of $G_{\phi,P}$ and $G_{P,P}$,
yields $P^{+}=\imath q_{0} \chi \phi^{+}+P^{+}_{N}$, where $\imath q_0 P^{+}_{N}=j^{+}_{N}$.
Here, we have assumed that the matter is homogeneous.
As it has been shown in Ref.~\!\cite{Soltani_flat_PRA_2010}, among the Green's functions $G_{\phi ,\phi }$,
$G_{\phi ,P }$ and $G_{P ,P}$, only $G_{\phi ,\phi }$ has contribution to the
partition function and consequently to the Casimir interaction. Therefore, hereafter we drop the subscript
$\phi$ from $G_{\phi ,\phi }$. Moreover, for both amplifying and dissipative media, $G_{\phi,\phi}$ is the Green's function
of Eq.~\!(\ref{varphi_relation}), therefore, the Casimir interaction between ideal conductors in the presence of
a dissipative or an amplifying slab (see Fig.~\!\ref{fig:schematic_1}) has the same mathematical form. The only
difference is in the sign of $\varepsilon_{\textrm{I}}(\omega)$.

TM and TE waves can be treated as two scalar fields separately which satisfy the following differential
equation $\left\{ \nabla ^2  - q_{0} ^2 \left[ {1 + \chi  (q_{0},\textbf{x} )} \right]\right\} \phi  = 0$.
For TE modes, $\phi$ and $\partial _z \phi$, and for TM waves, $\varepsilon (q_{0} )\phi$ and
$\partial _z \phi$ should be continuous on the left and also on the right surfaces of the slab, and in addition,
the Green's function, $G$, must satisfy the same BCs. Moreover, D and N BCs should be satisfied by TM and TE waves
on the surfaces of both conductors, respectively.

\subsection{Casimir interaction}\label{Casimir_interaction}

To obtain the Casimir interaction, one should calculate the partition function of the system
from the generating function. To this end, Wick rotation, $x_0 \rightarrow \imath \tau$, on
the time axis must be applied. After applying the above BCs on all fields, expressing D and N BCs in
terms of path-integral over the axillary fields \cite{Jalal_PRA_2006, Soltani_flat_PRA_2010,Soltani_rough_PRA_2010},
and integrating over all Gaussian fields, the partition function can be cast into
\begin{equation}
\ln {\cal Z}_{\textrm{TM(TE)}} 
= - \frac{1}{2}\ln \big[\det \Gamma_{\textrm{TM(TE)}} \big] ,
\label{partition_function}
\end{equation}
where
$\Gamma_{\textrm{TM(TE)}}$ is a second rank matrix with relevant
elements of 
$[\Gamma _{\textrm{TE}} ]_{\alpha\beta } =
\partial_{n} \partial^{'}_{n^{'}} G(x - y,z_{\alpha}  (x) ,z_{\beta} (y))$, 
$[\Gamma _{\textrm{TM}}]_{\alpha\beta } = G(x - y,z_{\alpha} (x) ,z_{\beta} (y))$,
where $\alpha, \beta = 1$ and 2, and $z_{1}  (x) = -H_{1}-\frac{b}{2}+h_{1}(x)$ and $z_{2}(x) = H_{2}+\frac{b}{2}+h_{2}(x)$. 
The Green's functions in the Fourier space can be read as
\begin{eqnarray}\label{Green_functions}
&& G_{\textrm{TM(TE)}} \big(\vec{q}, - H_1 -\frac{b}{2} , H_2 +\frac{b}{2} \big) = \nonumber\\
&& G_{\textrm{TM(TE)}} \big(\vec{q},H_2 +\frac{b}{2} , - H_1 - \frac{b}{2} \big) =
\nonumber\\
&& \hspace{+0.0cm} = \frac{{2\varepsilon_m Q_m e^{ - Q_1 (H_1  + H_2  - b)} }}{{(\varepsilon_m Q_1  + Q_m )^2
 \left[ { - (\Delta _{1m}^{\textrm{TM(TE)}} )^2  + e^{2Q_m b} } \right]}}, \nonumber\\
&&G_{\textrm{TM(TE)}} \big(\vec{q},( - 1)^j (H_j+\frac{b}{2}) ,( - 1)^j (H_j+\frac{b}{2}) \big) = \nonumber\\
&& \hspace{+0.0cm} =\frac{1}{2Q_1 } -\frac{{\Delta _{1m}^{\textrm{TM(TE)}} ( - 1 + e^{2Q_m b} )
 e^{Q_1 (b - 2H_j )} }}{{2Q_1 \left[ { - (\Delta _{1m}^{\textrm{TM(TE)}} )^2  + e^{2Q_2 b} } \right]}},
\end{eqnarray}
where $j=1,2$, $ \Delta _{1m}^{\textmd{TM}}  = \frac{{Q_m  - \varepsilon  (\dot \imath q_0 )\,Q_1 }}{{Q_m  + \varepsilon
(\dot \imath q_0 )\,Q_1 }}$, $\Delta _{1m}^{\textmd{TE}}  = \frac{{Q_m  - Q_1 }}{{Q_m  + Q_1 }}$,
$Q_{\zeta}  \equiv \dot \imath \sqrt {\varepsilon _{\zeta} (\imath q_{0} )q_{0} ^2  + q_1^2  + q_2^2 }$ with $\zeta= 1,m,2$
corresponds to regions $z<-b/2$, $|z|<b/2$, and $z>b/2$, respectively. The dielectric function of the slab is 
$\varepsilon_{m}(\imath q_{0}) =\varepsilon (\imath q_{0})$ while for $\zeta=1,2$, which corresponds to two vacuum space 
regions between slab and rough plates, $\varepsilon_{1}(\imath q_{0})=\varepsilon_{2}(\imath q_{0})=1$. 
For a general case where the dielectric slab has magnetic properties in addition to the electric ones, 
the definition of $\Delta$ must be adjusted accordingly. The general form of $\Delta$ is presented in Appendix~\!\ref{app1}.

Then, after performing the perturbative expansion up-to the second order with respect
to the deformation fields, $h_1(x)$ and $h_2 (x)$, the partition function is written as
\begin{eqnarray}\label{log_Part_func_1}
\ln \,{\cal Z}_{\gamma} = \ln \,{\cal Z}_{\gamma,0} + \ln \,{\cal Z}_{\gamma,2} ,
\end{eqnarray}
where the zeroth order of the partition function in the presence of the flat bounding conductors is
\begin{eqnarray}\label{log_Part_func_2}
&&\ln \,{\cal Z}_{\gamma,0} = - S \int \frac{d^3 \vec{q}}{(2 \pi)^3} \ln K_{\gamma}^0 (\vec{q}) ,
\end{eqnarray}
and the second order is
\begin{eqnarray}\label{log_Part_func_3}
&&\ln \,{\cal Z}_{\gamma,2} = \int d^3 xd^3 y \times \nonumber\\
&& \times \bigg\{ -\frac{1}{2} K_{\gamma}^{-} (x - y) 
[h_1 (x)h_2 (y) + h_2 (x)h_1 (y)] \nonumber\\
&& -\frac{1}{4} K_{\gamma}^{+} (x - y) 
\{ [h_1 (x)-h_1 (y)]^2 + [h_2 (x) - h_2 (y)]^2 \} \bigg\}, \nonumber\\
\end{eqnarray}
where the subscript, $\gamma= \textrm{TM or TE}$, stands for Dirichlet or Neumann BCs, respectively. It should be 
mentioned that the first order term of the logarithm of the partition function vanishes because we assumed that 
$\int h_i (x) d^3 x = 0$, where $i= 1$ and 2. The explicit forms of $K_{\gamma}^{-}$ and $K_{\gamma}^{+}$ for the 
geometry depicted in Fig.~\!\ref{fig:schematic_1} are presented in Appendices~\!\ref{kernels1} and \ref{kernels2}, 
and the explicit forms of $K_{\gamma}^{0}$ for different geometries can be seen in Sec.~\!\ref{sec_flat} and 
Appendices \ref{app1} and \ref{app2}. Using Eq.~\!(\ref{log_Part_func_1}), one can obtain the Casimir energy per 
unit area, $S$, as
\begin{equation}
E(H) = - \frac{\hbar c}{S L} \sum_{\gamma} \big[ \ln{\cal Z}_{\gamma} (H)
- \ln{\cal Z}_{\gamma} (H \rightarrow \infty)\big],
\label{E(H)}
\end{equation}
where $L$ is the overall Euclidean length in the temporal direction \cite{Golestanian_PRA_2003}. Then, Eq.~\!(\ref{E(H)})
is employed to study the Casimir interaction in different geometries in Secs.~\!\ref{sec_flat} and
\ref{sec_corrugation}.

\section{Flat Boundaries}\label{sec_flat}

\begin{figure}[t]
\includegraphics[width=0.96\columnwidth]{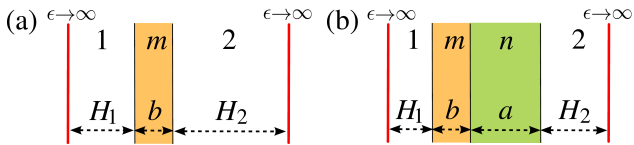}
\caption{(Color online) (a) Schematic picture of a dielectric slab in between two bounding ideal conductors
($\epsilon \rightarrow \infty$). The distance between the left conductor from the left
side of the slab is $H_1$, while the distance between the right conductor from the right
side of the slab is $H_2$, and the slab thickness is $b$. (b) Schematic picture of a double-layer
dielectric slab in between two bounding ideal conductors. The left dielectric layer thickness is $b$,
while the right dielectric layer thickness is $a$. The regions labeled by 1 and 2 are filled by the quantum
vacuum.}
\label{fig:schimatic_2}
\end{figure}

In this section, we study the normal Casimir interaction between either two flat ideal conductors 
($\varepsilon \rightarrow \infty$), or between a flat ideal conductor and a flat ideal permeable plate 
($\mu \rightarrow \infty$), in the presence of a flat dielectric or a flat double-layer dielectric slab
in between the ideal plates. Subsection \ref{subsec_counductor_conductor} is devoted to investigate the Casimir
force between two ideal conductors in the presence of a dissipative or an amplifying slab 
(see Fig.~\!\ref{fig:schimatic_2}a), or in the presence of a double-layer dielectric slab with one layer of 
dissipative and one layer of amplifying dielectrics (see Fig.~\!\ref{fig:schimatic_2}b), while in subsection 
\ref{subsec_counductor_permeable} we consider the Casimir force between an ideal conductor and an ideal permeable plate 
in the presence of either an absorptive or an amplifying or a double-layer dielectric slab.

To this end, we model the susceptibility of the amplifying slab by Lorentz model with gain and loss which is
one of the best choices to model the amplifying media \cite{Amooghorban_PRA_2011,Matloob_PRA_1997,Welsch_EPJST_2008}.
Linear gain occurs when the medium is pumped below the lasing threshold \cite{A. A. Zyablovsky}.
Therefore, we consider the following model for an amplifying medium
$\varepsilon_{\textrm{amp}} (\omega ) = 1 - \frac{{\omega_\textrm{p}^2 }}{{\omega _0^2  - \omega ^2  - \imath \gamma \omega }}$,
where $\omega_\textrm{p}= 0.75 \omega_0$, $\gamma= 10^{-3} \omega_0 $ and $\omega_0= 10^3$Hz \cite{Matloob_PRA_1997}.
It can be shown that for amplifying media, the imaginary part of the dielectric function is negative, i.e. 
$\varepsilon_{\textrm{I,amp}}(\omega)<0$. After applying Wick rotation, $\omega\rightarrow\imath q_{0}$, the above 
dielectric function is rewritten as
\begin{equation}
\varepsilon_{\textrm{amp}} (\imath q_{0}) = 1 - \frac{{\omega _\textrm{p}^2 }}{{\omega _0^2  + q_{0} ^2  + \gamma q_{0} }}.
\label{epsilon_amplifying}
\end{equation}
The dielectric function of the absorptive slab is also modeled here by Lorentz-oscillator model that after applying 
Wick rotation can be read as
\begin{equation}
\varepsilon_{\textrm{disp}} (\imath q_{0})=1+ \frac{{\omega _\textrm{p}^2 }}{{\omega _0^2 + q_{0} ^2 + \gamma q_{0} }}.
\label{epsilon_absorptive}
\end{equation}

\begin{figure*}[t]\begin{center}
\includegraphics[width=0.8\textwidth]{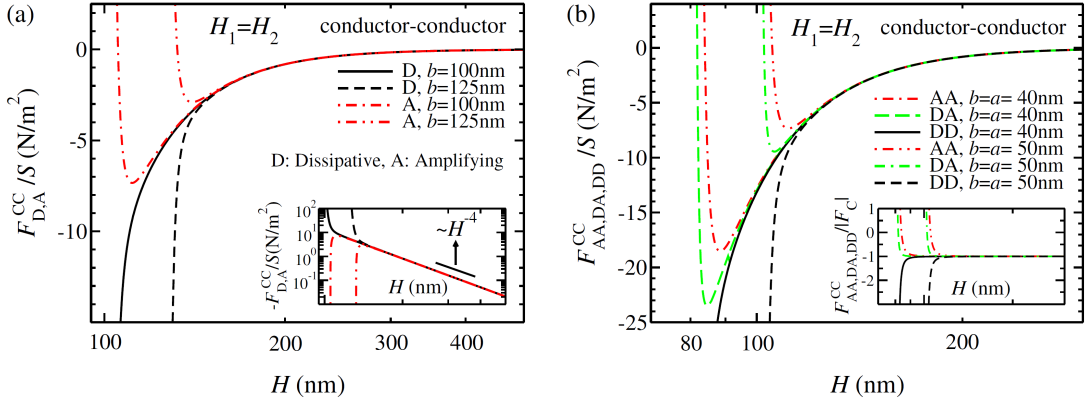}
\caption{(a) The force per unit area, $\frac{F_{m}^{\textrm{CC}}}{S}$, as a function of the distance between
the bounding conductors, $H= H_1 + H_2 + b$, with $H_1 = H_2$, in the presence of a single dielectric slab
between the bounding conductors, for two values of the dissipative (black solid and black dashed curves) and amplifying
(red dashed-dotted and red dashed-dotted-dotted curves) slab thicknesses $b= 100$ and $125$nm. The inset shows the log-log plot of
$-\frac{F_{m}^{\textrm{CC}}}{S}$ as a function of $H$. (b) The force per unit area, $\frac{F_{mn}^{\textrm{CC}}}{S}$,
as a function of the distance between the bounding conductors, $H= H_1 + H_2 + a+ b$, with $H_1 = H_2$,
in the presence of a double layer dielectric slab between the bounding conductors, for two values of the
dissipative and amplifying dielectric layers thicknesses $a=b= 40$ and $50$nm. The inset presents the normalized force
$F_{mn}^{\textrm{CC}}/ |F_C|$ as a function of $H$. Here, $m,n=\textrm{D}$ and A indicate the dissipative and
amplifying dielectrics, respectively.}
\label{fig:two_conductors}
\end{center}
\end{figure*}

\subsection{Conductor-Conductor}\label{subsec_counductor_conductor}
To obtain the Casimir interaction between two flat conductors we set the deformation fields for both conductors to
$h_1 (x) = h_2 (x) = 0$. Then the Casimir interaction energy per unit area for the system composed of a dielectric
slab in between two flat conductors, depicted in Fig.~\!\ref{fig:schimatic_2}a
reads as
\begin{eqnarray}
&&\hspace{-0.7cm}\frac{E_m^{\textrm{CC}}}{S} = \hbar c \int \frac{d^3 \vec{q}}{(2 \pi)^3}
\ln \bigg[ 1+ \frac{ {\mathcal{B}}_{m}^{\textrm{TM}} }{ {\mathcal{D}}_{m}^{\textrm{TM}} } \bigg]
+[(\textrm{TM}) \rightarrow (\textrm{TE})],
\label{energy_one_slab_two_conductors}
\end{eqnarray}
where the index $m={\textrm{D, A}}$ stands for a dissipative or an amplifying slab, respectively,
and ${\mathcal{D}}_{m}^{\textrm{TM}(\textrm{TE})}$ and ${\mathcal{B}}_{m}^{\textrm{TM}(\textrm{TE})}$ are functions of
$H_1$, $H_2$, $b$, $\varepsilon_1$, $\varepsilon_2$, $\varepsilon_m$ (permittivities of different layers),
$\mu_1$, $\mu_2$, $\mu_m$ (permeabilities of different layers),
$\vec{q}$, and $c$. The explicit
forms of ${\mathcal{D}}_{m}^{\textrm{TM}(\textrm{TE})}$ and ${\mathcal{B}}_{m}^{\textrm{TM}(\textrm{TE})}$ are
presented in Appendix~\!\ref{app1}. According to Eq.~\!(\ref{log_Part_func_2}), the kernel at the zeroth order with 
respect to the deformation fields, 
where $h_1 (x)=h_2 (x)=0$, is
$K_{\textrm{TM(TE)}}^0 = 1+ \frac{ {\mathcal{B}}_{m}^{\textrm{TM(TE)}} }{ {\mathcal{D}}_{m}^{\textrm{TM(TE)}} } $.

By using either path-integral formalism \cite{Ryder_book, Soltani_annals_2011, Soltani_flat_PRA_2010, Soltani_rough_PRA_2010} 
or transfer matrix method \cite{Podgornik_2003, Podgornik_2006, PRB_Jalal_2011} the Casimir interaction energy per unit area 
for the system composed of a double-layer dielectric slab in between two flat conductors, depicted in 
Fig.~\!\ref{fig:schimatic_2}b can be written as
\begin{eqnarray}
&&\hspace{-0.7cm}\frac{E_{mn}^{\textrm{CC}}}{S} = \hbar c \int \frac{d^3 \vec{q}}{(2 \pi)^3}
\ln \bigg[ 1+ \frac{ {\mathcal{B}}_{mn}^{\textrm{TM}} }{ {\mathcal{D}}_{mn}^{\textrm{TM}} } \bigg]
+[(\textrm{TM}) \rightarrow (\textrm{TE})],
\label{energy_one_slab_two_conductors}
\end{eqnarray}
where the indices $m,n={\textrm{D}}$ and A stand for a dissipative and an amplifying slab, respectively,
and ${\mathcal{D}}_{mn}^{\textrm{TM}(\textrm{TE})}$ and ${\mathcal{B}}_{mn}^{\textrm{TM}(\textrm{TE})}$ 
are functions of $a$ and $\mu_n$ in addition to $H_1$, $H_2$, $b$, $\varepsilon_1$, $\varepsilon_2$, $\varepsilon_m$, 
$\varepsilon_n$, $\mu_1$, $\mu_2$, $\mu_m$, $\mu_n$, $\vec{q}$ and $c$. The explicit forms of 
${\mathcal{D}}_{mn}^{\textrm{TM}(\textrm{TE})}$ and ${\mathcal{B}}_{mn}^{\textrm{TM}(\textrm{TE})}$
are written in Appendix~\!\ref{app1}.

To show the effect of the presence of the dielectric slab in between two conductors on the Casimir interaction
in the system depicted in Fig.~\!\ref{fig:schimatic_2}a, the Casimir force per unit area that the right
conductor experiences at fixed $H_1$ and at fixed $b$ can be obtained as
\begin{equation}
\frac{F_{m}^{\textrm{CC}}}{S} = - \frac{\partial E_{m}^{\textrm{CC}}  }{\partial H_2} \bigg|_{H_1, b}.
\label{force_on_the_right_conductor}
\end{equation}
Similar procedure is performed to calculate the Casimir force on the right conductor in the presence
of the double-layer dielectrics (see Fig.~\!\ref{fig:schimatic_2}b) as
\begin{equation}
\frac{F_{mn}^{\textrm{CC}}}{S} = - \frac{\partial E_{mn}^{\textrm{CC}}  }{\partial H_2} \bigg|_{H_1, a, b}.
\label{force_on_the_right_conductor_double-layer}
\end{equation}

To compare the difference between $F_{\textrm{D}}^{\textrm{CC}}$, the Casimir force in the presence of the 
dissipative slab, and $F_{\textrm{A}}^{\textrm{CC}}$, the Casimir force in the presence of an amplifying slab, 
in Fig.~\!\ref{fig:two_conductors}a the force per unit area has been plotted as a function of the distance 
between the bounding conductors, $H= H_1 + H_2 + b$, with $H_1 = H_2$, for two values of the dissipative 
(black solid and black dashed curves) and amplifying (red dashed-dotted and red dashed-dotted-dotted curves) slab 
thicknesses $b= 100$, and $125$nm. As it is clear, the force in the presence of the dissipative dielectric slab 
is attractive for the whole distance range between the bounding conductors, while in the presence of the 
amplifying slab, for small distances the force is repulsive and as the distance increases the force decreases 
and beyond a certain distance its sign changes and it becomes attractive. Therefore, there is an equilibrium 
distance, $H_{\textrm{equil}}$, where the value of the force is zero. This equilibrium distance is a function 
of the amplifying slab thickness, its dielectric function and the distance between the bounding conductors, $H$. 
By increasing the thickness of the amplifying slab, the location of the equilibrium point is shifted to the 
larger distances. The inset shows the log-log plot of $-\frac{F_{m}^{\textrm{CC}}}{S}$ 
as a function of $H$, which approximately reveals that when $H_2 > b/8$ the force scales as $H^{-4}$. In 
Fig.~\!\ref{fig:two_conductors}b, the Casimir force per unit area, $\frac{F_{mn}^{\textrm{CC}}}{S}$, in the presence of 
a double-layer slab between two bounding conductors depicted in Fig.~\!\ref{fig:schimatic_2}b, has been plotted as a 
function of $H$, with $H_1 = H_2$, for two values of the dissipative and amplifying slabs thicknesses $a, b= 40$
and $50$nm. Here, $m, n=\textrm{D}$ and A indicate the dissipative and amplifying dielectrics, respectively.
This force is attractive when both layers are dissipative (DD). When one of the layer is dissipative and the other 
is amplifying (DA), at small distances the force is repulsive. By increasing $H$, the force becomes attractive and it gets a 
minimum and then it grows. For DA slab the system has a mechanical equilibrium state with the equilibrium distance
$H_{\textrm{DA,equil}}$. When both layers of the slab are amplifying (AA) the equilibrium point is shifted to the larger 
distances and $H_{\textrm{AA,equil}} > H_{\textrm{DA,equil}}$. The inset represents the normalized force, 
$F_{mn}^{\textrm{CC}}/ |F_C|$, as a function of $H$. Again when approximately $H_2 > (a+b)/8$, the force scales as $H^{-4}$.

\begin{figure}[t]
\includegraphics[width=0.96\columnwidth]{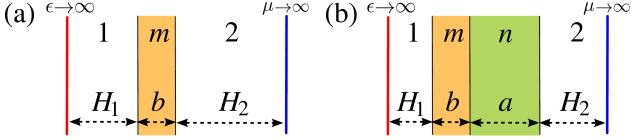}
\caption{(Color online) (a) Schematic picture of a dielectric slab in between two bounding plates, one is
flat ideal conductor ($\epsilon \rightarrow \infty$) while the other is a flat ideal permeable plate ($\mu \rightarrow \infty$).
The distance between the ideal conductor from the left side of the slab is $H_1$, while the distance
between the permeable plate from the right side of the slab is $H_2$ and the slab thickness is $b$.
(b) Schematic picture of a double-layer dielectric slab in between an ideal conductor and an ideal permeable
plate. The left dielectric layer thickness is $b$, while the right dielectric layer thickness is $a$. The regions
labeled by 1 and 2 are filled by the quantum vacuum.}
\label{fig:schematic_3}
\end{figure}

\begin{figure*}[t]\begin{center}
\includegraphics[width=0.8\textwidth]{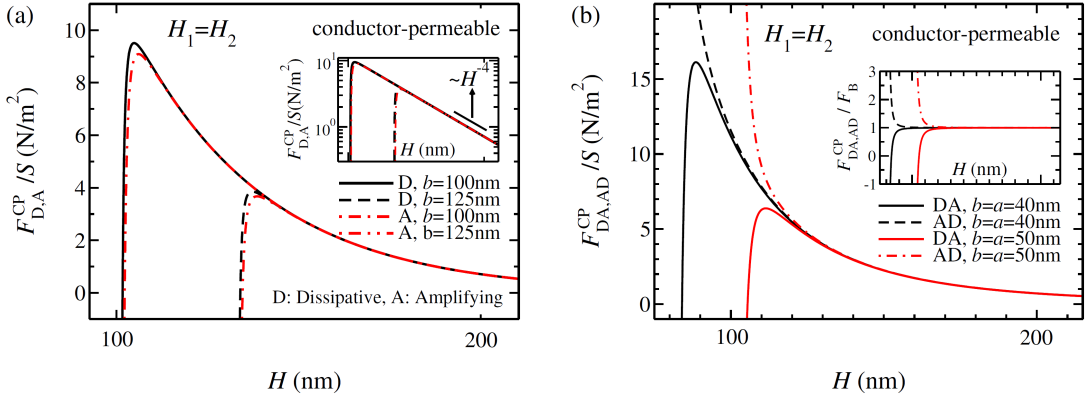}
\caption{(a) The force per unit area, $\frac{F_{m}^{\textrm{CP}}}{S}$, as a function of the distance between
the bounding conductor and the permeable plate, $H= H_1 + H_2 + b$, with $H_1 = H_2$, in the presence of a single
dielectric slab between the bounding ideal plates, for two values of the dissipative (black solid and black dashed
curves) and amplifying (red dashed-dotted and red dashed-dotted-dotted curves) slab thicknesses $b= 100$ and $125$nm.
The inset shows the log-log plot of $\frac{F_{m}^{\textrm{CP}}}{S}$ as a function of $H$. (b) The force per
unit area, $\frac{F_{mn}^{\textrm{CP}}}{S}$, as a function of the distance between the bounding ideal plates,
$H= H_1 + H_2 + a+ b$, with $H_1 = H_2$, in the presence of a double layer dielectric slab between the bounding
ideal plates, for two values of the dissipative and amplifying slab thicknesses $a= b= 40$ and $50$nm. The
inset presents the normalized force, $F_{mn}^{\textrm{CP}}/ F_B$, where $F_B =- \frac{7}{8} F_C$, as a function of
$H$. Here, $m,n=\textrm{D}$ and A indicate the dissipative and amplifying dielectrics, respectively.}
\label{fig:conductor-permeable}
\end{center}
\end{figure*}

\subsection{Conductor-Permeable}\label{subsec_counductor_permeable}

To observe how much the characteristics of the bounding plates can affect the Casimir interaction, we consider
a system composed of a single dielectric slab, or a double-layer dielectric slab in between a flat ideal conductor 
in the left side and a flat ideal permeable plate in the right side of the system, depicted in Fig.~\!\ref{fig:schematic_3}a 
and b, respectively. As in 1974 Boyer showed \cite{Boyer}, a flat conductor and a flat permeable plate immersed in the
quantum vacuum and face each other at a distance $H$ in the absence of dielectric slab, experience a repulsive force
$F_{B}= -\frac{7}{8} F_{C}$ \cite{kiani_PRA_2012, Hushwater, Jalal_EPL_2015}. Here, our aim is to investigate the
effect of the dissipative and amplifying slabs on the Boyer repulsive force, $F_B$. Using either path-integral technique
\cite{Ryder_book, Soltani_annals_2011, Soltani_flat_PRA_2010, Soltani_rough_PRA_2010} or transfer matrix formalism
\cite{Podgornik_2003, Podgornik_2006, PRB_Jalal_2011} the Casimir
interaction energy per unit area for a system composed of a single dielectric slab in between an ideal conductor
and an ideal permeable plate (see Fig.~\!\ref{fig:schematic_3}a) reads as
\begin{eqnarray}
&&\hspace{-0.7cm}\frac{E_m^{\textrm{CP}}}{S} =\hbar c \int \frac{d^3 \vec{q}}{(2 \pi)^3}
\ln \bigg[ 1+ \frac{ {\mathcal{I}}_{m}^{\textrm{TM}} }{ {\mathcal{J}}_{m}^{\textrm{TM}} } \bigg]
+[(\textrm{TM}) \rightarrow (\textrm{TE})],
\label{energy_one_slab_one_conductor_permeable}
\end{eqnarray}
and for a system composed of a double-layer dielectric slab in between an ideal conductor and an ideal permeable
plate (see Fig.~\!\ref{fig:schematic_3}b) as
\begin{eqnarray}
&&\hspace{-0.7cm}\frac{E_{mn}^{\textrm{CP}}}{S} = \hbar c \int \frac{d^3 \vec{q}}{(2 \pi)^3}
\ln \bigg[ 1+ \frac{ {\mathcal{I}}_{mn}^{\textrm{TM}} }{ {\mathcal{J}}_{mn}^{\textrm{TM}} } \bigg]
+[(\textrm{TM}) \rightarrow (\textrm{TE})],
\label{energy_one_slab_conductor_permeable}
\end{eqnarray}
where again the indices $m, n={\textrm{D}}$ and A stand for a dissipative and an amplifying slab, respectively,
${\mathcal{I}}_{m}^{\textrm{TM}(\textrm{TE})}$ and ${\mathcal{J}}_{m}^{\textrm{TM}(\textrm{TE})}$ are functions
of $H_1$, $H_2$, $b$, $\varepsilon_1$, $\varepsilon_2$, $\varepsilon_m$, $\mu_1$, $\mu_2$, $\mu_m$, $\vec{q}$ and $c$, and
${\mathcal{I}}_{mn}^{\textrm{TM}(\textrm{TE})}$ and ${\mathcal{J}}_{mn}^{\textrm{TM}(\textrm{TE})}$ are functions of
the above parameters in addition to $a$ and $\mu_n$.
The explicit forms of ${\mathcal{I}}_{m}^{\textrm{TM}(\textrm{TE})}$, ${\mathcal{J}}_{m}^{\textrm{TM}(\textrm{TE})}$,
${\mathcal{I}}_{mn}^{\textrm{TM}(\textrm{TE})}$ and ${\mathcal{J}}_{mn}^{\textrm{TM}(\textrm{TE})}$ are presented in
Appendix~\!\ref{app2}.

To compare the effects of a dissipative or an amplifying slab between two ideal flat plates, one conductive and the other
permeable, on the Casimir interaction, the force per unit area that the permeable plate on the right side experiences can 
be calculated at the fixed $H_1$ and at fixed $b$ as
\begin{equation}
\frac{F_{m}^{\textrm{CP}}}{S} = - \frac{\partial E_{m}^{\textrm{CP}}  }{\partial H_2} \bigg|_{H_1, b}.
\label{force_on_the_right_permeable}
\end{equation}
Similar procedure can be performed to calculate the Casimir force on the right side permeable plate in the presence
of the double-layer dielectric slab in between, for the fixed values of $H_1, b$ and $a$, as
\begin{equation}
\frac{F_{mn}^{\textrm{CP}}}{S} = - \frac{\partial E_{mn}^{\textrm{CP}}  }{\partial H_2} \bigg|_{H_1, a, b}.
\label{force_on_the_right_permeable_double-layer}
\end{equation}

In Fig.~\!\ref{fig:conductor-permeable}a, the force per unit area, $\frac{F_{m}^{\textrm{CP}}}{S}$, has been plotted
as a function of the distance between the bounding conductor and the permeable plate, $H= H_1 + H_2 + b$, with
$H_1 = H_2$, in the presence of a single dielectric slab, for two values of the dissipative (black solid and black dashed
curves) and amplifying (red dashed-dotted and red dashed-dotted-dotted curves) slabs thicknesses $b= 100$, and $125$nm.
The inset shows the log-log plot of $\frac{F_{m}^{\textrm{CP}}}{S}$ as a function of $H$. The overall behavior of
the forces due to the presence of the dissipative slab and due to the presence of the amplifying slab are similar to
each other. At very small distances the bounding plates attract each other. By increasing the distance, the force
increases and it gets its zero value at the distance $H_{\textrm{D(A),equil}}$, where $H_{\textrm{D,equil}} <
H_{\textrm{A,equil}}$.

To show the effect of the asymmetry in the system due to the presence of conducting and permeable plates, on the Casimir force
in the presence of the double-layer dielectric slab, the slab is located at the middle of the cavity by setting
$H_1= H_2$, and we choose $b=a$. The Casimir force is then calculated for the geometries i) conductor-vacuum-dissipative
dielectric layer-amplifying dielectric layer-vacuum-permeable plate (DA), ii) conductor-vacuum-amplifying dielectric layer-dissipative
dielectric layer-vacuum-permeable plate (AD). Quite interestingly, the force is different for the geometries i and ii. To present
this interesting phenomenon, in Fig.~\!\ref{fig:conductor-permeable}b, the force per unit area, $\frac{F_{mn}^{\textrm{CP}}}{S}$,
has been plotted as a function of the distance between the bounding plates, $H= H_1 + H_2 + a+ b$, for DA and AD geometries, 
for two values of the dissipative and amplifying slabs thicknesses $a=b= 40$ and $50$nm. The force is purely repulsive for
AD geometry, while for DA, the cross over from attractive to repulsive force occurs when $H$ increases. By increasing the 
thickness of the double-layer slab for DA geometry, the location of the equilibrium point is shifted to the larger distances.
It should be mentioned that in both Figs.~\!\ref{fig:conductor-permeable}a and b, the equilibrium states for the position 
of the right ideal plate are unstable. The inset presents the normalized force, $F_{mn}^{\textrm{CP}}/ F_B$, as a function of $H$.

\section{Corrugated Boundaries: Conductor-Conductor}\label{sec_corrugation}
\subsection{Normal Interaction Between a Flat and a Corrugated Conductors} \label{subsec_normal}

\begin{figure*}[t]\begin{center}
\includegraphics[width=0.8\textwidth]{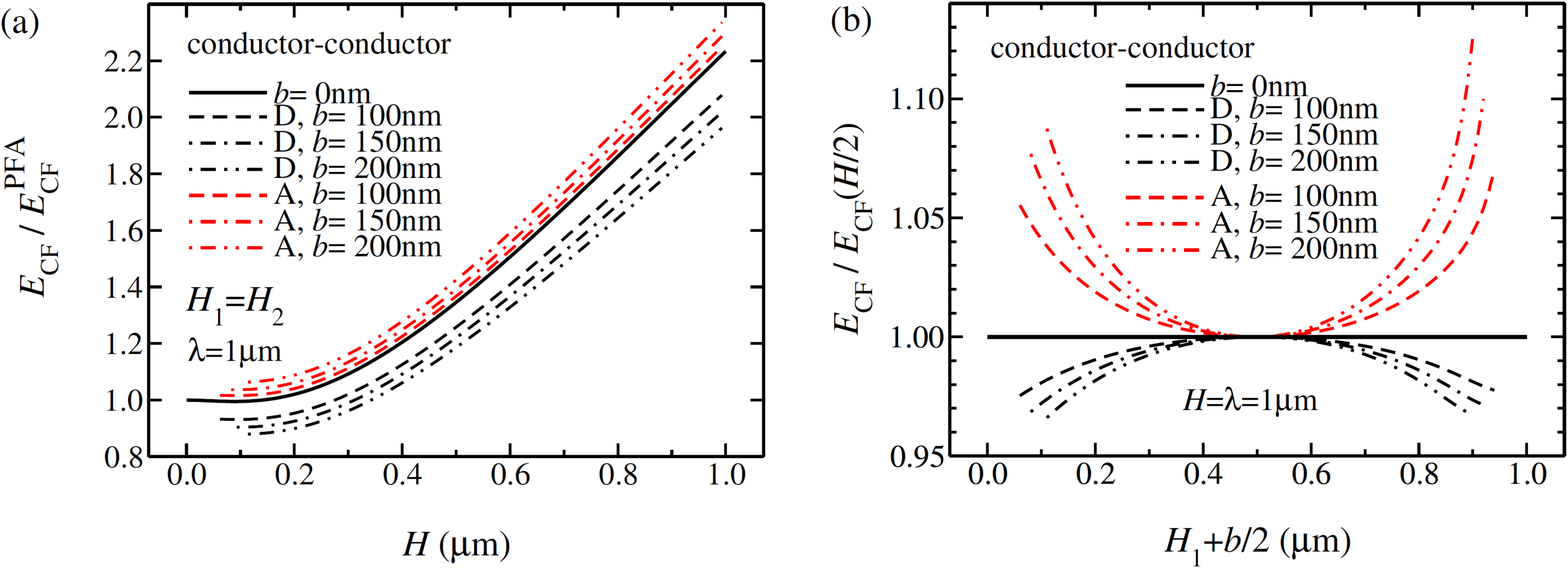}
\caption{(a) The normalized Casimir energy, $E_{\textrm{CF}} / E_{\textrm{CF}}^{\textrm{PFA}} $, as a function of 
$H= H_1 + H_2 + b$ with $H_1 = H_2$ and $\lambda = 1 \mu$m in the presence of a dissipative (D, black curves) or 
an amplifying (A, red curves) dielectric slab for various values of the slab thicknesses $b=0, 100, 150$ and $200$nm. 
The black solid curve shows the normalized energy in the absence of dielectric slab. (b) The normalized Casimir 
energy, $\frac{E_{\textrm{CF}}}{ E_{\textrm{CF}} (H/2) }$, as a function of $H_1 + b/2$ which is the distance between 
the left conductor and the middle of the absorptive (black curves) and amplifying slabs (red curves), for fixed values 
of $\lambda = H= 1 ~\! \mu \textrm{m}$ and different values of $b$ the same as those of panel (a). Here, 
$E_{\textrm{CF}} (H/2)$ is the value of $E_{\textrm{CF}}$ when the center of the slab is located at the distance $H/2$ 
from each conductors. i.e. $H_1 = H_2$.}
\label{fig:Ecf}
\end{center}
\end{figure*}

To investigate the effect of corrugation on the normal Casimir interaction between two conductors in the presence of
a dissipative or an amplifying slab, we set the deformation fields on the conductors $h_1 (x)=0$ and
$h_2 (x) = a \cos [2\pi x_1 /\lambda ]$ (see Fig.~\!\ref{fig:schematic_1}). Then the Casimir energy
due to the corrugation on one of the conductors, labeled by 2, is obtained as
\begin{equation}\label{Ecf_1}
E_{\textrm{CF}} = \frac{\hbar c}{A L} \sum_{\gamma} \big[ \ln{\cal Z}_{\gamma,2} (H) -
\ln{\cal Z}_{\gamma,2} (H\rightarrow \infty) \big] |_{h_1 (x) =0},
\end{equation}
where index CF stands for corrugated-flat, and sum is over $\gamma= \textrm{TM and TE}$. This energy can then be cast into
\begin{equation}\label{Ecf_2}
{E_{\textrm{CF}} = -\frac{\pi^2 a^2 \hbar c}{240 H^5}
\big[ K_{\textrm{TM}}^{+ \textrm{reg}} (\frac{H}{\lambda }) + K_{\textrm{TE}}^{+ \textrm{reg}} (\frac{H}{\lambda }) \big],}
\end{equation}
where according to Eq.~\!(\ref{Ecf_1}), the regular part of the TM and TE kernels is defined as
$ K_{\textrm{TM(TE)}}^{+ \textrm{reg}}= K_{\textrm{TM(TE)}}^{+} - \mathop{\textrm{Lim}}\limits_{H_1  \to \infty }
K_{\textrm{TM(TE)}}^{+}$. In Eq.~\!(\ref{Ecf_2}), $ K_{\textrm{TM(TE)}}^{+} (q)$ is the Fourier transformation of
$K_{\textrm{TM(TE)}}^{+}(x)$ at $\vec{q}=(0,\frac{2\pi}{\lambda},0)$. The explicit forms of
$K_{\textrm{TM}}^{+} (\frac{H}{\lambda })$ and $K_{\textrm{TE}}^{+} (\frac{H}{\lambda })$
are presented in Appendices~\!\ref{kernels1} and \ref{kernels2}. In Fig.~\!\ref{fig:Ecf}a the normalized contribution 
of the corrugation to the Casimir energy, $\frac{E_{\textrm{CF}}}{ E_{\textrm{CF}}^{\textrm{PFA}} }$, has been plotted 
as a function of $H= H_1 + H_2 + b$ with $H_1 = H_2$ and $\lambda = 1 \mu$m in the presence of a dissipative (D, black curves) 
or an amplifying (A, red curves) dielectric slab with thicknesses $b=0, 100, 150$ and $200$nm. The black solid curve shows 
the normalized energy in the absence of dielectric slab. Quite interestingly, the presence of an amplifying slab enhances 
the Casimir interaction due to the roughness while the interaction is weakened by the presence of the absorptive slab. This 
effect is more pronounced by increasing the slab thickness. The dielectric functions of the amplifying and dissipative slabs 
are the same as of Eqs.~\!(\ref{epsilon_amplifying}) and (\ref{epsilon_absorptive}), respectively.

The superscript PFA in $E_{\textrm{CF}}^{\textrm{PFA}}$ stands for proximity force approximation. PFA is used when $a$,
which is the corrugation amplitude, is much smaller than the other length scales in the system such as $H$ and $\lambda$.
In the proximity force approximation, the surface elements of a curved surface around each point are simply replaced
by surface elements parallel to the plane of $(x_1,x_2)$ at the same point \cite{Boordag_Book,Lambretch_PFA}. The Casimir
energy in a system composed of a corrugated and a flat conductors in PFA is
$E_{\textrm{CF}}^{\textrm{PFA}} = - \frac{a^2}{4} \frac{\partial F_{C}}{\partial H}$.

To study the effect of the location of the dielectric slab on the Casimir interaction between the bounding plates,
in Fig.~\!\ref{fig:Ecf}b, the normalized Casimir energy, $\frac{E_{\textrm{CF}}}{ E_{\textrm{CF}} (H/2) }$, has been
shown as a function of $H_1 + b/2$ which is the distance between the left conductor and the middle of the dissipative
slab (black curves) or the middle of the amplifying slab (red curves), for fixed values of 
$\lambda = H= 1 ~\! \mu \textrm{m}$ and different values of $b$ which are the same as those of panel (a). Here,
$E_{\textrm{CF}} (H/2)$ is the value of $E_{\textrm{CF}}$ when the center of the slab is located at the distance
$H/2$ from each conductors. As it can be seen, when the amplifying slab is closer to the rough conductor in the 
right side, this effect is more pronounced.

\begin{figure*}[t]\begin{center}
\includegraphics[width=0.8\textwidth]{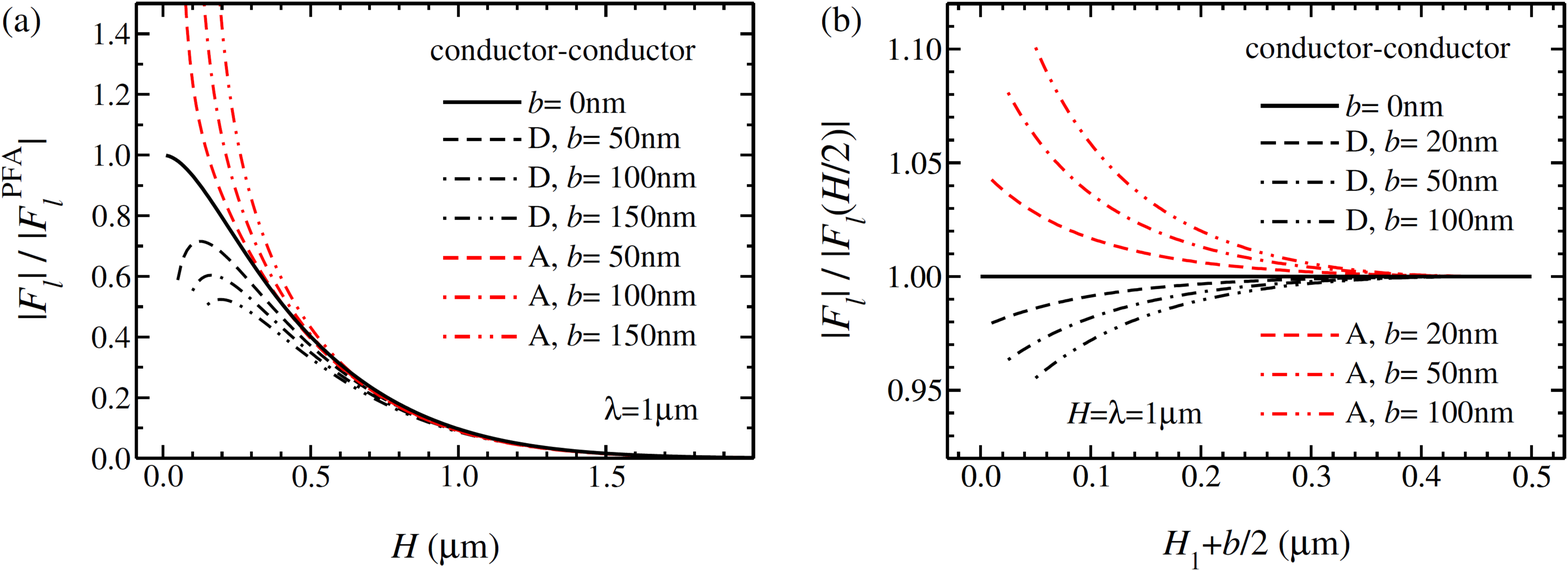}
\caption{(a) The normalized amplitude of the lateral Casimir force, $|F_l|/|F_l^{\textrm{PFA}}|$, between rough metallic
plates (depicted in Fig.~\!\ref{fig:schematic_1}) as a function of the mean distance between corrugated conductors,
$H= H_1 + H_2 + b$, for fixed corrugation wavelength $\lambda=1~\!\mu \textrm{m}$, and various
values of dissipative slab thicknesses $b=0, 50, 100$ and $150~\!\textrm{nm}$ (black curves, from top to bottom)
and amplifying slab thicknesses (red curves, from bottom to top) the same as of the dissipative one. Here, the
center of the slab is fixed at the middle distance between bounding ideal conductors. (b) The normalized amplitude of the
lateral Casimir force, $|F_l|/|F_{l} (H/2)|$, as a function of $H_1 + b/2$ which is the distance between the
left conductor and the middle of the dissipative (black curves) or amplifying slab (red curves),
for fixed values of $\lambda = H= 1 ~\! \mu \textrm{m}$ and different values of the dielectric slab thicknesses 
$b= 0, 20, 50$ and $100$nm. Here, $|F_{l} (H/2)|$ is the amplitude of the lateral force when the center of the slab 
is located at the distance $H/2$ from each conductors.}
\label{fig:lateral_force}
\end{center}
\end{figure*}

\subsection{Lateral Force Between Two Corrugated Conductors}\label{subsec_lateral}

Two corrugated conductors, immersed in the quantum vacuum, can experience the lateral Casimir force due to the translational
symmetry breaking \cite{Golestanian_PRA_2003,Jalal_PRA_2006,Zakeri_PRA_2012} in addition to the normal Casimir force. In this
section, we compare the effect of the presence of the absorptive or amplifying dielectric slabs on the lateral Casimir 
force between the bounding corrugated conductors depicted in Fig.~\!\ref{fig:schematic_1}. The lateral Casimir force is calculated 
by $F_l  = - \frac{\partial E}{\partial l} $, where $E$ is the Casimir energy in Eq.~\!(\ref{E(H)}). By setting the corrugation
fields on the rough conductors depicted in Fig.~\!\ref{fig:schematic_1} as
$h_1 (x)=a \cos [2\pi x_1 /\lambda ]$ and $h_2 (x) = a \cos [2\pi (x_1 +l) /\lambda ]$, the lateral Casimir force is then
obtained as
\begin{equation}\label{eq:lateral_force}
F_l  = \frac{2 \pi \hbar c a^2 }{ \lambda H^5 } \sin (\frac{{2\pi l}}{\lambda })[ K_{\textrm{TM}}^{-} (\frac{H}{\lambda })
+ K_{\textrm{TE}}^{-} (\frac{H}{\lambda })],
\end{equation}
where $K_{\textrm{TM}}^{-} (q)$ and $K_{\textrm{TE}}^{-} (q)$ are the Fourier transformations of $K_{\textrm{TM}}^{-}(x)$ 
and $K_{\textrm{TE}}^{-}(x)$ at $\vec{q}=(0,\frac{2\pi}{\lambda},0)$, respectively, and their explicit forms can be seen 
in Appendices~\!\ref{kernels1} and \ref{kernels2}.

In Fig.~\!\ref{fig:lateral_force}(a), we have plotted the normalized amplitude of the lateral Casimir force, 
$|F_l|/|F_l^{\textrm{PFA}}|$, between two rough metallic plates depicted in Fig.~\!\ref{fig:schematic_1} as a function of 
$H$ (the mean separation distance between corrugated conductors) for fixed corrugation wavelength, $\lambda=1~\!\mu \textrm{m}$
and various values of the absorptive (black curves) amplifying (red curves) slabs thicknesses $b=0, 50, 100$ and $150~\!\textrm{nm}$. 
Here, the center of the slab is fixed at the middle distance between plates. $|F_l^{\textrm{PFA}}|$ is the magnitude of the 
lateral Casimir force for the same geometry in PFA \cite{Boordag_Book,Lambretch_PFA}, which is
$|F_l^{\textrm{PFA}}| = \frac{\pi a^2}{\lambda} \frac{\partial F_C}{\partial H}$.
Comparing the value of the normalized lateral Casimir force in the presence of the dissipative and amplifying slabs yields
that at small distances, the lateral Casimir force in the presence of an amplifying slab is stronger than that of absorptive
dielectric slab with the same thickness, whereas by increasing the distance, in both cases, the values of the normalized force
approach to the values of the force in the absence of the dielectric slab. Moreover, by increasing the thickness 
of an amplifying slab, the lateral Casimir force increases whereas for the absorptive one decreases.

We have also considered the effect of the location of the center of the dissipative or amplifying slabs, $H_1 + b/2$,
on the lateral Casimir force. To this end, in Fig.~\!\ref{fig:lateral_force}b, the normalized amplitude of the lateral Casimir
force, $|F_l|/|F_{l} ( H/2)|$, has been shown as a function of $H_1 + b/2$ which is the mean distance between the left conductor
and the middle of the dissipative (black curves) or amplifying (red curves) slab, for fixed values of
$\lambda = H= 1 ~\! \mu \textrm{m}$ and different values of $b=0, 20, 50$ and 100nm. $|F_{l} (H/2)|$ is the amplitude of 
the lateral force when the center of the slab is fixed at the middle of the cavity. The dielectric functions of the amplifying 
and absorptive slabs are chosen the same as in Eqs.~\!(\ref{epsilon_amplifying}) and (\ref{epsilon_absorptive}), respectively. 
The sinusoidal behavior of the force does not alter by changing $H_1$, while its amplitude alters. Moreover, if the 
amplifying slab comes closer to each conductor, the amplitude of the lateral Casimir force increases. In contrast to the 
amplifying slab, the amplitude of the lateral Casimir force decreases when the dissipative dielectrics comes closer to 
one of the bounding conductors. By increasing the thickness of the slab, this effect is more pronounced.

\section{Conclusion}\label{sec_conclusion}

Using either path-integral formalism or transfer matrix method, we investigated the normal Casimir interaction
between flat ideal plates in the presence of a dissipative or an amplifying slab, or a double-layer dielectric slab. The normal force 
between flat ideal conductors is non-monotonic due to the presence of the amplifying slab and the system has a stable equilibrium 
state, while the force is attractive and is weakened by intervening the absorptive dielectric slab in the cavity. By replacing 
the right flat conductor by an ideal permeable plate, the overall behaviors of the Casimir force in the presence of the absorptive 
slab and in the presence of the amplifying slab are the same, and for both cases, the force is non-monotonic and the system has an 
unstable equilibrium state. Quite interestingly, the Casimir force is non-monotonic in the presence of a double-layer dielectric 
slab in the DA geometry while it is purely repulsive for the AD geometry.
Then employing the path-integral technique, we calculated the correction to the normal Casimir interaction due to the corrugation
on one of the bounding conductors, and also the lateral Casimir force has been obtained due to the roughness on both bounding
conductors in the presence of the absorptive or amplifying slab. While the presence of the amplifying slab enhances both normal and
lateral Casimir interactions between the bounding conductors due to the roughness, the presence of the absorptive slab weakens both
normal and lateral interactions compared to those of a cavity contains only vacuum between the bounding conductors.
We also showed that both normal and lateral Casimir interactions depend on the distance of the slab from the bounding conductors.
By approaching the amplifying slab to one of the conductors, both normal and lateral Casimir interactions enhance compared to
the Casimir interactions in a cavity which contains only vacuum between the bounding conductors. If instead, the dissipative dielectric
slab is brought closer to one of the bounding conductors both normal and lateral Casimir interactions are weakened.
The reason for the difference in the behavior of the Casimir interactions can be understood if one looks at the characteristics of the
vacuum fluctuations of the EM field near the slab. The dissipative medium decreases the vacuum fluctuations of the EM field whereas the
amplifying medium increases these fluctuations \cite{Welsch_EPJST_2008,Welsch_Book}. The method that we employed in this paper can
be used to study the dynamic Casimir effect \cite{Jalal_PRA_2006,Jalal_PRA_2007} and the Casimir interaction at finite
temperature \cite{PRB_Jalal_2011, Jalal_PRA_2011} in the presence of absorptive or amplifying slabs.

\begin{acknowledgments}
J.S. acknowledges support from the Academy of Finland through its Centers
of Excellence Program (2012-2017) under Project No. 915804.
\end{acknowledgments}

\appendix
\section{{Fourier transformation of $K_{\textrm{TM}}^{+(-)} (x)$}}\label{kernels1}

The Fourier transformations of the kernels for TM waves at $(0, \frac{2\pi}{\lambda}, 0 )$ are 
\begin{widetext}
\begin{eqnarray}
K_{\textrm{TM}}^{+} (0,\frac{2\pi}{\lambda},0)&=& \int \frac{d^3 \vec{q} }{(2\pi)^3} 
\bigg[ + F_1 (\vec q,H_1 ,H_1 ) Q_1^2 G _{\textrm{TM}} \big(\vec q + 
\frac{2\pi}{\lambda}\hat i,H_2+\frac{b}{2} ,H_2+\frac{b}{2} \big)
+ F_1 (\vec q,H_1 ,H_1 )F_5 (\vec q + \frac{2\pi}{\lambda}\hat{i},H_2 ,H_2 )\nonumber \\
&& \hspace{+1.6cm} + F_3 (\vec q,H_1 ,H_1 )F_3 (\vec q + \frac{2\pi}{\lambda}\hat i,H_2 ,H_2 ) \bigg],\nonumber\\
K_{\textrm{TM}}^{-} (0,\frac{2\pi}{\lambda},0) &=& \int \frac{d^3 \vec{q} }{(2\pi)^3} 
\bigg[ +F_2 (\vec q,H_1 ,H_1 )  Q_1^2 G_{\textrm{TM}} (\vec q + \frac{2\pi}{\lambda}\hat i,H_2+\frac{b}{2} ,H_2+\frac{b}{2} )
+ F_4 (\vec{q},H_1 ,H_1 )F_6 (\vec{q} + \frac{2\pi}{\lambda}\hat{i},H_2 ,H_2 ) \nonumber\\
&& \hspace{+1.6cm} + F_2 (\vec{q},H_1 ,H_1 )F_2 (\vec{q} + \frac{2\pi}{\lambda}\hat{i},H_2 ,H_2 )\bigg],
\label{kernels_TM}
\end{eqnarray}
where
\begin{eqnarray}
F_1 (\vec{q},H_i ,H_i ) &=& 
\frac{{G_{\textrm{TM}} \big(\vec{q},(-1)^i(H_i+\frac{b}{2}) ,(-1)^i(H_i+\frac{b}{2}) \big)}}
{{N(\vec{q},H_1 ,H_2 )}}, \nonumber\\
F_2 (\vec{q},H_i ,H_i ) &=& 
\frac{{G_{\textrm{TM}} \big(\vec{q},(-1)^i(H_i+\frac{b}{2}) ,(-1)^i(H_i+\frac{b}{2}) \big)}}
{{N(\vec{q},H_1 ,H_2 )}}
Q_1 G_{\textrm{TM}}\big(\vec{q},-H_1-\frac{b}{2} ,H_2+\frac{b}{2} \big),\nonumber\\
F_3 (\vec{q},H_1 ,H_2 ) &=& 
\frac{{G_{\textrm{TM}}(\vec{q},-H_1-\frac{b}{2} ,H_2+\frac{b}{2} \big)}}{{N(\vec{q},H_1 ,H_2 )}}
Q_1 G_{\textrm{TM}}\big(\vec{q},-H_1-\frac{b}{2} ,H_2+\frac{b}{2} \big),\nonumber\\
F_4 (\vec{q},H_1 ,H_2 ) &=& 
\frac{{G_{\textrm{TM}}\big(\vec{q},-H_1-\frac{b}{2} ,H_2+\frac{b}{2} \big)}}{{N(\vec{q},H_1 ,H_2 )}}, 
\nonumber\\
F_5 (\vec{q},H_i ,H_i ) 
&=& \frac{{G_{\textrm{TM}}\big(\vec{q},(-1)^i(H_i+\frac{b}{2} ,(-1)^i(H_i+\frac{b}{2}) \big)}}
{{N(\vec{q},H_1 ,H_2 )}}
Q_1^2 G_{\textrm{TM}}\big(\vec{q},-H_1-\frac{b}{2} ,H_2+\frac{b}{2} \big), \nonumber\\
F_6 (\vec{q},H_1 ,H_2 ) &=& 
\frac{{G_{\textrm{TM}}\big(\vec{q},-H_1-\frac{b}{2} ,H_2+\frac{b}{2} \big)}}{{N(\vec{q},H_1 ,H_2 )}}
Q_1^2 G_{\textrm{TM}}\big(\vec{q},-H_1-\frac{b}{2} ,H_2+ \frac{b}{2} \big),
\end{eqnarray}
and
\begin{eqnarray}
N(\vec{q},H_1,H_2) &=& G_{\textrm{TM}} 
\big(\vec q,-H_1-\frac{b}{2},-H_1-\frac{b}{2}\big)
G_{\textrm{TM}} \big(\vec q,H_2+\frac{b}{2},H_2+\frac{b}{2}\big)
-G_{\textrm{TM}}^2 \big(\vec q,-H_1-\frac{b}{2},H_2+\frac{b}{2}\big).
\end{eqnarray}
\end{widetext}

\section{{Fourier transformation of $K_{\textrm{TE}}^{-(+)} (x)$}}\label{kernels2}
The Fourier transformations of the kernels for TE waves at $(0, \frac{2\pi}{\lambda}, 0 )$ are 
\begin{widetext}
\begin{eqnarray}
K_{\textrm{TE}}^{+}  (0,\frac{2\pi}{\lambda} ,0) &=& \int \frac{d^3 \vec{q} }{(2\pi)^3} 
\bigg\{ {\cal F}_1 (\vec{q},H_1 ,H_1 ) ~ Q_1^2 ~ g \big(\vec{q} + \frac{2\pi}{\lambda}\hat{i},H_2 ,H_2 \big)
+{\cal F}_1 (\vec{q},H_1 ,H_1 ){\cal F}_5 (\vec{q} + \frac{2\pi}{\lambda}\hat{i},H_2 ,H_2 )\nonumber \\
&& \hspace{-2.0cm} + {\cal F}_3 (\vec{q},H_1 ,H_2 ){\cal F}_3 (\vec{q} + \frac{2\pi}{\lambda}\hat i,H_1 ,H_2 ) 
+ \big(\frac{2\pi}{\lambda} \big)^2 
\bigg[F_1 (\vec{q},H_1 ,H_1 ) \big(q_1  + \frac{2\pi}{\lambda} \big)^2 
G_{\textrm{TE}} \big(\vec{q} + \frac{2\pi}{\lambda}\hat{i},H_2+\frac{b}{2} ,H_2+\frac{b}{2} \big) \nonumber\\ 
&& \hspace{-2.0cm} + {\cal F}_1 (\vec{q},H_1 ,H_1 ) \big(q_1  + \frac{2\pi}{\lambda} \big)^2 
{\cal F}_7 (\vec{q} + \frac{2\pi}{\lambda}\hat{i},H_2 ,H_2 )
+ q_1 \big(q_1  + \frac{2\pi}{\lambda} \big) {\cal F}_9 (\vec q,H_1 ,H_2 ) 
{\cal F}_9 (\vec q + \frac{2\pi}{\lambda}\hat{i},H_1 ,H_2 ) \bigg] \nonumber\\ 
&& \hspace{-2.0cm} + \frac{2\pi}{\lambda} 
\bigg[ F_1 (\vec{q},H_1 ,H_1 ) \big(q_1 + \frac{2\pi}{\lambda} \big) g\big(\vec{q} + \frac{2\pi}{\lambda}\hat i,H_2 ,H_2 \big) 
+ {\cal F}_1 (\vec{q},H_1 ,H_1 ) \big(q_1 + \frac{2\pi}{\lambda} \big) 
{\cal F}_{11} (\vec{q} + \frac{2\pi}{\lambda}\hat{i},H_2 ,H_2 ) \nonumber \\ 
&& \hspace{-2.0cm} +
{\cal F}_2 (\vec{q},H_1 ,H_1 ) \big(q_1 + \frac{2\pi}{\lambda} \big) 
{\cal F}_{10} (\vec{q} + \frac{2\pi}{\lambda}\hat i,H_2 ,H_2 ) \bigg]
\bigg\}, \nonumber \\
K_{\textrm{TE}}^{-}  (0,\frac{2\pi}{\lambda} ,0) &=& \int \frac{d^3 \vec{q} }{(2\pi)^3}
\bigg\{ {\cal F}_4 (\vec{q},H_1 ,H_2 ) ~ Q_1^2 ~ g\big(\vec{q} + \frac{2\pi}{\lambda}\hat{i},H_2 ,H_2 \big)
+ {\cal F}_4 (\vec{q},H_1 ,H_2 ){\cal F}_6 (\vec{q} + \frac{2\pi}{\lambda}\hat{i},H_1 ,H_2 ) \nonumber \\ 
&& \hspace{-2.0cm} + {\cal F}_2 (\vec{q},H_1 ,H_1 ){\cal F}_2 (\vec{q} + \frac{2\pi}{\lambda}\hat{i},H_2 ,H_2 ) 
+ \big( \frac{2\pi}{\lambda} \big)^2 \bigg[{\cal F}_4 (\vec q,H_1 ,H_2 ) \big(q_1  + \frac{2\pi}{\lambda} \big)^2 
G_{\textrm{TE}}(\vec{q} + \frac{2\pi}{\lambda}\hat{i},H_2+\frac{b}{2} ,H_2+\frac{b}{2} )\nonumber \\ 
&& \hspace{-2.0cm} + {\cal F}_4 (\vec{q},H_1 ,H_2 ) \big(q_1  + \frac{2\pi}{\lambda} \big)^2 
{\cal F}_8 (\vec q + \frac{2\pi}{\lambda}\hat i,H_1 ,H_2 ) 
+ q_1 \big(q_1  + \frac{2\pi}{\lambda} \big) {\cal F}_{10} (\vec{q},H_1 ,H_1 )
{\cal F}_{10} (\vec{q} + \frac{2\pi}{\lambda}\hat{i},H_1 ,H_2 ) \bigg]\nonumber \\ 
&& \hspace{-2.0cm} + 2\frac{2\pi}{\lambda} 
\bigg[ {\cal F}_4 (\vec{q},H_1 ,H_2 ) \big(q_1  + \frac{2\pi}{\lambda} \big) 
g\big(\vec{q} + \frac{2\pi}{\lambda}\hat{i},H_1 ,H_2 \big)
+{\cal F}_4 (\vec{q},H_1 ,H_2 ) \big(q_1  + \frac{2\pi}{\lambda} \big)
{\cal F}_{12} (\vec{q} + \frac{2\pi}{\lambda}\hat{i},H_1 ,H_2 )\nonumber \\ 
&& \hspace{-2.0cm} + {\cal F}_2 (\vec{q},H_1 ,H_1 ) \big(q_1  + \frac{2\pi}{\lambda} \big)
{\cal F}_{10} (\vec{q} + \frac{2\pi}{\lambda}\hat{i},H_2 ,H_2 )\bigg]
\bigg\} ,
\end{eqnarray}
\end{widetext}
where
\begin{eqnarray}
{\cal F}_1 (\vec{q},H_i ,H_i ) &=& \frac{{g(\vec{q},H_i ,H_i )}}{{{\cal N}(\vec{q},H_1 ,H_2 )}}, \nonumber\\
{\cal F}_2 (\vec{q},H_i ,H_i ) &=& \frac{{g(\vec{q},H_i ,H_i )}}{{{\cal N}(\vec{q},H_1 ,H_2 )}} Q_1g(\vec{q},H_1 ,H_2 ), \nonumber\\
{\cal F}_3 (\vec{q},H_1 ,H_2 ) &=& \frac{{g(\vec{q},H_1,H_2 )}}{{{\cal N}(\vec{q},H_1 ,H_2 )}}Q_1 g(\vec{q},H_1 ,H_2 ), \nonumber\\
{\cal F}_4 (\vec{q},H_1 ,H_2 ) &=& \frac{{g(\vec{q},H_1 ,H_2 )}}{{{\cal N}(\vec{q},H_1 ,H_2 )}} , \nonumber\\
{\cal F}_5 (\vec{q},H_i ,H_i ) &=& \frac{{g(\vec{q},H_i ,H_i )}}{{{\cal N}(\vec{q},H_1 ,H_2 )}}Q_1^2 g(\vec{q},H_1 ,H_2 ) , \nonumber\\
{\cal F}_6 (\vec{q},H_1 ,H_2 ) &=& \frac{{g(\vec{q},H_1 ,H_2 )}}{{{\cal N}(\vec{q},H_1 ,H_2 )}}Q_1^2 g(\vec{q},H_1 ,H_2 ) , \nonumber\\
{\cal F}_7 (\vec{q},H_i ,H_i ) &=& \frac{{g(\vec{q},H_i ,H_i )}}{{{\cal N}(\vec{q},H_1 ,H_2 )}}Q_1^2\nonumber\\
&& \times G_{\textrm{TE}}(\vec{q},-H_1-\frac{b}{2},H_2+\frac{b}{2} ) , \nonumber\\
{\cal F}_8 (\vec{q},H_1 ,H_2 ) &=& \frac{{g(\vec{q},H_1 ,H_2 )}}{{{\cal N}(\vec{q},H_1 ,H_2 )}}Q_1^2 \nonumber\\
&& \times G_{\textrm{TE}}(\vec{q},-H_1-\frac{b}{2},H_2+\frac{b}{2} ) ,
\end{eqnarray}
\begin{eqnarray}
{\cal F}_9 (\vec{q},H_1 ,H_2 ) &=& \frac{{g(\vec{q},H_1 ,H_2 )}}{{{\cal N}(\vec{q},H_1 ,H_2 )}}Q_1 \nonumber\\
&& \times G_{\textrm{TE}}(\vec{q},-H_1-\frac{b}{2},H_2+\frac{b}{2} ) , \nonumber\\
{\cal F}_{10} (\vec{q},H_i ,H_i ) &=& \frac{{g(\vec{q},H_i ,H_i )}}{{{\cal N}(\vec{q},H_1 ,H_2 )}}Q_1 \nonumber\\
&& \times G_{\textrm{TE}}(\vec{q},-H_1-\frac{b}{2},H_2+\frac{b}{2} ) , \nonumber\\
{\cal F}_{11} (\vec{q},H_i ,H_i ) &=& \frac{{g(\vec{q},H_i ,H_i )}}{{{\cal N}(\vec{q},H_1 ,H_2 )}}Q_1 \nonumber\\
&& \hspace{-2.0cm} \times  G_{\textrm{TE}}(\vec{q},-H_1-\frac{b}{2},H_2+\frac{b}{2} )Q_1 g(\vec{q},H_1 ,H_2 ) , \nonumber\\
{\cal F}_{12} (\vec{q},H_1 ,H_2 ) &=& \frac{{g(\vec{q},H_1 ,H_2 )}}{{{\cal N}(\vec{q},H_1 ,H_2 )}}Q_1 \nonumber\\
&& \hspace{-2.0cm} \times G_{\textrm{TE}}(\vec{q},-H_1-\frac{b}{2},H_2+\frac{b}{2} )Q_1 g(\vec{q},H_1 ,H_2 ) ,
\end{eqnarray}
and
\begin{equation}
g(\vec{q},H_i,H_j ) =  Q_1^2 ~\! G_{\textrm{TE}} \big( \vec{q},(-1)^i(H_i+\frac{b}{2}) ,(-1)^j(H_j+\frac{b}{2}) \big),
\end{equation}
\begin{equation}
{\cal N}(\vec{q},H_1 ,H_2 ) = g(\vec{q},H_1 ,H_1 )g(\vec{q},H_2 ,H_2 ) - g^2 (\vec{q},H_1 ,H_2 ).
\end{equation}

\section{Conductor-Conductor}\label{app1}
The definitions of ${\mathcal{B}}_{m}^{\textrm{TM}}$, ${\mathcal{B}}_{m}^{\textrm{TE}}$, ${\mathcal{D}}_{m}^{\textrm{TM}}$
and ${\mathcal{D}}_{m}^{\textrm{TE}}$ are

\begin{widetext}
\begin{eqnarray}
{\mathcal{B}}_{m}^{\textrm{TM}} &=& - \Delta_{m1}^{\textrm{TM}} e^{-2 Q_1 H_1} +
\Delta_{2m}^{\textrm{TM}} e^{-2 Q_2 H_2} 
- \Delta_{m1}^{\textrm{TM}} \Delta_{2m}^{\textrm{TM}} e^{-2 (Q_1 H_1 + Q_2 H_2) }
- \Delta_{2m}^{\textrm{TM}} e^{-2 (Q_1 H_1 + Q_m b) }
+ \Delta_{m1}^{\textrm{TM}}  e^{-2 (Q_2 H_2 + Q_m b) }\nonumber\\
&&-  e^{-2 (Q_1 H_1 + Q_2 H_2 + Q_m b) }, \nonumber\\
{\mathcal{D}}_{m}^{\textrm{TM(TE)}} &=& 1 + \Delta_{m1}^{\textrm{TM(TE)}} \Delta_{2m}^{\textrm{TM(TE)}}
e^{-2 Q_m b } , \nonumber\\
{\mathcal{B}}_{m}^{\textrm{TE}} &=& + \Delta_{m1}^{\textrm{TE}} e^{-2 Q_1 H_1} -
\Delta_{2m}^{\textrm{TM}} e^{-2 Q_2 H_2} - \Delta_{m1}^{\textrm{TE}} \Delta_{2m}^{\textrm{TE}} e^{-2 (Q_1 H_1 + Q_2 H_2) }
+ \Delta_{2m}^{\textrm{TE}} e^{-2 (Q_1 H_1 + Q_m b) } - \Delta_{m1}^{\textrm{TE}}  e^{-2 (Q_2 H_2 + Q_m b) } \nonumber\\
&&-  e^{-2 (Q_1 H_1 + Q_2 H_2 + Q_m b) }, 
\label{D_one_slab_two_conductors_1}
\end{eqnarray}
where
\begin{equation}
\Delta_{ij}^{\textrm{TM}} =
\frac{ \varepsilon_i (\imath q_0) Q_j (\imath q_0) - \varepsilon_j (\imath q_0) Q_i (\imath q_0) }
{ \varepsilon_i (\imath q_0) Q_j (\imath q_0) + \varepsilon_j (\imath q_0) Q_i (\imath q_0) } ,
\hspace{+0.5cm}\Delta_{ij}^{\textrm{TE}} =
\frac{ \mu_i (\imath q_0) Q_j (\imath q_0) - \mu_j (\imath q_0) Q_i (\imath q_0) }
{ \mu_i (\imath q_0) Q_j (\imath q_0) +  \mu_j (\imath q_0) Q_i (\imath q_0) },
\label{Delta}
\end{equation}
with the layer labeled by $j$ is located at the left-hand side of the layer labeled by $i$.
$Q_i^2 = {\bf q}^2 + \frac{\varepsilon_i (\imath q_0) \mu_i (\imath q_0) q_0^2}{c^2}$, ${\bf q}^2 = q_1^2 + q_2^2 $ and $c$
is the speed of light in the vacuum.
Functions ${\mathcal{B}}_{mn}^{\textrm{TM}}$, ${\mathcal{B}}_{mn}^{\textrm{TE}}$, ${\mathcal{D}}_{mn}^{\textrm{TM}}$
and ${\mathcal{D}}_{mn}^{\textrm{TE}}$ are

\begin{eqnarray}
{\mathcal{B}}_{mn}^{\textrm{TM}} &=&
+ \Delta_{2n}^{\textrm{TM}} e^{-2 Q_2 H_2} +
\Delta_{nm}^{\textrm{TM}} e^{-2 (Q_n a + Q_2 H_2)} - \Delta_{m1}^{\textrm{TM}} e^{-2 Q_1 H_1 }
- \Delta_{2n}^{\textrm{TM}} \Delta_{m1}^{\textrm{TM}} e^{-2 (Q_1 H_1 + Q_2 H_2) }  \nonumber\\
&& - \Delta_{2n}^{\textrm{TM}} \Delta_{nm}^{\textrm{TM}} \Delta_{m1}^{\textrm{TM}}
e^{-2 (Q_n a + Q_1 H_1) }- \Delta_{nm}^{\textrm{TM}} \Delta_{m1}^{\textrm{TM}}  e^{-2 (Q_1 H_1 + Q_2 H_2 + Q_n a) } 
- \Delta_{nm}^{\textrm{TM}} e^{-2 (Q_m b + Q_1 H_1 ) }  \nonumber\\
&& + \Delta_{2n}^{\textrm{TM}} \Delta_{nm}^{\textrm{TM}} \Delta_{m1}^{\textrm{TM}}
e^{-2 (Q_m b + Q_2 H_2) }
- \Delta_{2n}^{\textrm{TM}} \Delta_{nm}^{\textrm{TM}}  e^{-2 (Q_1 H_1 + Q_2 H_2 + Q_m b) }
- \Delta_{2n}^{\textrm{TM}} e^{-2 (Q_n a + Q_m b + Q_1 H_1 ) }  \nonumber\\
&& + \Delta_{m1}^{\textrm{TM}} e^{-2 (Q_n a + Q_m b + Q_2 H_2 ) }
-  e^{-2 ( Q_1 H_1 + Q_2 H_2 + Q_n a + Q_m b ) }, \nonumber\\
{\mathcal{D}}_{mn}^{\textrm{TM(TE)}} &=&
1 + \Delta_{2n}^{\textrm{TM(TE)}} \Delta_{nm}^{\textrm{TM(TE)}}
e^{-2 Q_n a } + \Delta_{m1}^{\textrm{TM(TE)}} \Delta_{nm}^{\textrm{TM(TE)}} e^{-2 Q_m b }
+ \Delta_{2n}^{\textrm{TM(TE)}} \Delta_{m1}^{\textrm{TM(TE)}} e^{-2 ( Q_n a + Q_m b ) }, \nonumber\\
{\mathcal{B}}_{mn}^{\textrm{TE}} &=& - \Delta_{2n}^{\textrm{TE}} e^{-2 Q_2 H_2} -
\Delta_{nm}^{\textrm{TE}} e^{-2 (Q_n a + Q_2 H_2)} 
+ \Delta_{m1}^{\textrm{TE}} e^{-2 Q_1 H_1 }
- \Delta_{2n}^{\textrm{TE}} \Delta_{m1}^{\textrm{TE}} e^{-2 (Q_1 H_1 + Q_2 H_2) }  \nonumber\\
&&+ \Delta_{2n}^{\textrm{TE}} \Delta_{nm}^{\textrm{TE}} \Delta_{m1}^{\textrm{TE}}
e^{-2 (Q_n a + Q_1 H_1) } - \Delta_{nm}^{\textrm{TE}} \Delta_{m1}^{\textrm{TE}}  e^{-2 (Q_1 H_1 + Q_2 H_2 + Q_n a) } 
+ \Delta_{nm}^{\textrm{TE}} e^{-2 (Q_m b + Q_1 H_1 ) }  \nonumber\\
&&- \Delta_{2n}^{\textrm{TE}} \Delta_{nm}^{\textrm{TE}} \Delta_{m1}^{\textrm{TE}}
e^{-2 (Q_m b + Q_2 H_2) } - \Delta_{2n}^{\textrm{TE}} \Delta_{nm}^{\textrm{TE}}  e^{-2 (Q_1 H_1 + Q_2 H_2 + Q_m b) }
+ \Delta_{2n}^{\textrm{TE}} e^{-2 (Q_n a + Q_m b + Q_1 H_1 ) }  \nonumber\\
&& - \Delta_{m1}^{\textrm{TE}} e^{-2 (Q_n a + Q_m b + Q_2 H_2 ) } -  e^{-2 ( Q_1 H_1 + Q_2 H_2 + Q_n a + Q_m b ) }.
\label{B_two_slab_two_conductors_1}
\end{eqnarray}

\section{Conductor-Permeable}\label{app2}
Functions ${\mathcal{I}}_{m}^{\textrm{TM}}$, ${\mathcal{I}}_{m}^{\textrm{TE}}$, ${\mathcal{J}}_{m}^{\textrm{TM}}$,
${\mathcal{J}}_{m}^{\textrm{TE}}$, ${\mathcal{I}}_{mn}^{\textrm{TM}}$, ${\mathcal{I}}_{mn}^{\textrm{TE}}$, 
${\mathcal{J}}_{mn}^{\textrm{TM}}$ and ${\mathcal{J}}_{mn}^{\textrm{TE}}$ are 

\begin{eqnarray}
{\mathcal{I}}_{m}^{\textrm{TM}} &=& - \Delta_{m1}^{\textrm{TM}} e^{-2 Q_1 H_1} -
\Delta_{2m}^{\textrm{TM}} e^{-2 Q_2 H_2} + \Delta_{m1}^{\textrm{TM}} \Delta_{2m}^{\textrm{TM}} e^{-2 (Q_1 H_1 + Q_2 H_2) }
- \Delta_{2m}^{\textrm{TM}} e^{-2 (Q_1 H_1 + Q_m b) } 
- \Delta_{m1}^{\textrm{TM}}  e^{-2 (Q_2 H_2 + Q_m b) } \nonumber\\
&&+  e^{-2 (Q_1 H_1 + Q_2 H_2 + Q_m b) }, \nonumber\\
{\mathcal{J}}_{m}^{\textrm{TM(TE)}} &=& 1 + \Delta_{m1}^{\textrm{TM(TE)}} \Delta_{2m}^{\textrm{TM(TE)}}
e^{-2 Q_m b } , \nonumber\\
{\mathcal{I}}_{m}^{\textrm{TE}} &=& + \Delta_{m1}^{\textrm{TE}} e^{-2 Q_1 H_1} +
\Delta_{2m}^{\textrm{TM}} e^{-2 Q_2 H_2} + \Delta_{m1}^{\textrm{TE}} \Delta_{2m}^{\textrm{TE}} e^{-2 (Q_1 H_1 + Q_2 H_2) }
+ \Delta_{2m}^{\textrm{TE}} e^{-2 (Q_1 H_1 + Q_m b) } + \Delta_{m1}^{\textrm{TE}}  e^{-2 (Q_2 H_2 + Q_m b) } \nonumber\\
&& +  e^{-2 (Q_1 H_1 + Q_2 H_2 + Q_m b) }, \nonumber\\
{\mathcal{I}}_{mn}^{\textrm{TM}} &=& 
- \Delta_{2n}^{\textrm{TM}} e^{-2 Q_2 H_2} - \Delta_{nm}^{\textrm{TM}} e^{-2 (Q_n a + Q_2 H_2)} 
- \Delta_{m1}^{\textrm{TM}} e^{-2 Q_1 H_1 }
+ \Delta_{2n}^{\textrm{TM}} \Delta_{m1}^{\textrm{TM}} e^{-2 (Q_1 H_1 + Q_2 H_2) }  \nonumber\\
&& - \Delta_{2n}^{\textrm{TM}} \Delta_{nm}^{\textrm{TM}} \Delta_{m1}^{\textrm{TM}}
e^{-2 (Q_n a + Q_1 H_1) } + \Delta_{nm}^{\textrm{TM}} \Delta_{m1}^{\textrm{TM}}  e^{-2 (Q_1 H_1 + Q_2 H_2 + Q_n a) } 
- \Delta_{nm}^{\textrm{TM}} e^{-2 (Q_m b + Q_1 H_1 ) } \nonumber\\
&& - \Delta_{2n}^{\textrm{TM}} \Delta_{nm}^{\textrm{TM}} \Delta_{m1}^{\textrm{TM}}
e^{-2 (Q_m b + Q_2 H_2) } + \Delta_{2n}^{\textrm{TM}} \Delta_{nm}^{\textrm{TM}}  e^{-2 (Q_1 H_1 + Q_2 H_2 + Q_m b) } 
- \Delta_{2n}^{\textrm{TM}} e^{-2 (Q_n a + Q_m b + Q_1 H_1 ) }  \nonumber\\
&& - \Delta_{m1}^{\textrm{TM}} e^{-2 (Q_n a + Q_m b + Q_2 H_2 ) } +  e^{-2 ( Q_1 H_1 + Q_2 H_2 + Q_n a + Q_m b ) }, \nonumber\\
{\mathcal{J}}_{mn}^{\textrm{TM(TE)}} &=&
1 + \Delta_{2n}^{\textrm{TM(TE)}} \Delta_{nm}^{\textrm{TM(TE)}} e^{-2 Q_n a } 
+ \Delta_{m1}^{\textrm{TM(TE)}} \Delta_{nm}^{\textrm{TM(TE)}} e^{-2 Q_m b } 
+ \Delta_{2n}^{\textrm{TM(TE)}} \Delta_{m1}^{\textrm{TM(TE)}} e^{-2 ( Q_n a + Q_m b ) }, \nonumber\\
{\mathcal{I}}_{mn}^{\textrm{TE}} &=&
+ \Delta_{2n}^{\textrm{TE}} e^{-2 Q_2 H_2} +
\Delta_{nm}^{\textrm{TE}} e^{-2 (Q_n a + Q_2 H_2)} + \Delta_{m1}^{\textrm{TE}} e^{-2 Q_1 H_1 }
+ \Delta_{2n}^{\textrm{TE}} \Delta_{m1}^{\textrm{TE}} e^{-2 (Q_1 H_1 + Q_2 H_2) }  \nonumber\\
&& + \Delta_{2n}^{\textrm{TE}} \Delta_{nm}^{\textrm{TE}} \Delta_{m1}^{\textrm{TE}}
e^{-2 (Q_n a + Q_1 H_1) } 
+ \Delta_{nm}^{\textrm{TE}} \Delta_{m1}^{\textrm{TE}}  e^{-2 (Q_1 H_1 + Q_2 H_2 + Q_n a) } 
+ \Delta_{nm}^{\textrm{TE}} e^{-2 (Q_m b + Q_1 H_1 ) }  \nonumber\\
&& + \Delta_{2n}^{\textrm{TE}} \Delta_{nm}^{\textrm{TE}} \Delta_{m1}^{\textrm{TE}}
e^{-2 (Q_m b + Q_2 H_2) } 
+ \Delta_{2n}^{\textrm{TE}} \Delta_{nm}^{\textrm{TE}}  e^{-2 (Q_1 H_1 + Q_2 H_2 + Q_m b) } 
+ \Delta_{2n}^{\textrm{TE}} e^{-2 (Q_n a + Q_m b + Q_1 H_1 ) }  \nonumber\\
&& + \Delta_{m1}^{\textrm{TE}} e^{-2 (Q_n a + Q_m b + Q_2 H_2 ) }  
+  e^{-2 ( Q_1 H_1 + Q_2 H_2 + Q_n a + Q_m b ) }.
\label{I_one_slab_two_conductors_3}
\end{eqnarray}
\end{widetext}

\end{document}